\documentclass[prd,twocolumn,superscriptaddress,showkeys,notitlepage]{revtex4-2}

\usepackage[english]{babel}
\usepackage{xcolor}
\usepackage{mathtools}
\usepackage{pdfsync}

\usepackage{amssymb}
\usepackage{amsmath}
\usepackage{amsthm}
\usepackage{amsfonts}

\usepackage{dsfont}
\usepackage{amssymb}
\usepackage{mathrsfs}
\RequirePackage{mathptmx} 

\catcode`\@=11
\usepackage{empheq}
\usepackage{tcolorbox}

\usepackage{color,colordvi}
 
\usepackage{slashed}

\usepackage[colorlinks=true,
linkcolor=purple,
anchorcolor=magenta,
citecolor=blue
]{hyperref}

\usepackage{xcolor}

\def\d{\hbox{{d}\kern-.20em\hbox{l}}}

\def \matrix #1 {\left(\begin{array}{cc} #1 \end{array}\right)}

\def\II{\hbox{{1}\kern-.25em\hbox{l}}}

\newcommand{\BMP}{\scriptscriptstyle{\mathrm{BMP}}} 
\newcommand{\BMJ}{\scriptscriptstyle{\mathrm{KM}}}

\def \t {{\widehat t}}

\def \mm {\widehat{m}^2}
\def \Pperp {|\widehat{P}_\perp|^2}

\def \Dw {\mathcal{D}_w}

 \definecolor{myblue}{rgb}{.8, .8, 1}
 \definecolor{shadecolor}{rgb}{0.92,0.9,0.9}
 \definecolor{light-gray}{rgb}{0.827451,0.827451,0.827451}

\def \Li {\text{Li\,}}

\def\inbar{\,\vrule height1.5ex width.4pt depth0pt}
\def\IC{\relax\hbox{$\inbar\kern-.3em{\rm C}$}}
\def\IZ{\relax{\hbox{\cmss Z\kern-.4em Z}}}
\def\IR{{\hbox{{\rm I}\kern-.2em\hbox{\rm R}}}}

\def\IP{{\hbox{{\rm I}\kern-.2em\hbox{\rm P}}}}
\def\II{\hbox{{1}\kern-.25em\hbox{l}}}

\begin{document}


\title{
\hfill \small\textmd{TUM-HEP-1551/25}\\
\hfill \small\textmd{DESY 24--223}
\\
Kinematic power corrections to DVCS to twist-six accuracy}

\author{V. M. Braun}
   \affiliation{Institut f\"ur Theoretische Physik, Universit\"at
   Regensburg, D-93040 Regensburg, Germany}

\author{Yao Ji}
\affiliation{School of Science and Engineering, The Chinese University of Hong Kong, 518172 Shenzhen, China}
\affiliation{Physics Department T31, Technische Universit{\"a}t M{\"u}nchen, D-85748 Garsching, Germany}

  \author{A. N. Manashov}
\affiliation{II. Institut f\"ur Theoretische Physik, Universit\"at Hamburg
   D-22761 Hamburg, Germany}
\affiliation{Institut f\"ur Theoretische Physik, Universit\"at
   Regensburg, D-93040 Regensburg, Germany}

\begin{abstract}
We calculate  $(\sqrt{-t}/Q)^k $ and $(m/Q)^k$ power corrections with $k\le 4$, 
where $m$ is the target mass and $t$ is the momentum transfer, to several key observables in 
Deeply Virtual Compton Scattering (DVCS).
We find that the power expansion is well convergent up to $|t|/Q^2\lesssim 1/4$  
for most of the observables, but is naturally organized in terms of 
$1/(Q^2+t)$ rather than the nominal hard scale $1/Q^2$. 
We also argue that target mass corrections remain under control and do not endanger QCD 
factorization for coherent DVCS on nuclei.     
These results remove an important source of uncertainties 
due to the frame dependence and violation of electromagnetic Ward identities 
in the QCD predictions for the DVCS amplitudes in the leading-twist approximation.

\end{abstract}


\keywords{DVCS,  GPD, higher twist}

\maketitle


                                          \section{Introduction}\label{sec:intro}


Studies of the deeply-virtual Compton scattering (DVCS) play an important role in the quest for the three-dimensional ``tomographic'' imaging of the proton and 
light nuclei. This reaction gives access to the generalized parton distributions
(GPDs)~\cite{Muller:1994ses,Ji:1996nm,Radyushkin:1997ki}  that encode the information on the transverse position of quarks and gluons in
the proton in dependence on their longitudinal momentum.  This is an active research topic and a major science goal for the
planned  Electron-Ion Collider (EIC) \cite{AbdulKhalek:2021gbh,AbdulKhalek:2022erw}. 
The QCD description of the DVCS is based on collinear factorization. At leading power, the complete next-to-leading-order (NLO) results are
available since long ago~\cite{Ji:1998xh,Belitsky:1999hf,Belitsky:1998gc,Noritzsch:2003un}. A lot of effort is put into extending this
description to NNLO~\cite{Kumericki:2006xx,Kumericki:2007sa,Braun:2017cih,Braun:2020yib,Braun:2021grd,Gao:2021iqq,
Braun:2022byg,Braun:2022bpn,Ji:2023eni,Ji:2023xzk,Schoenleber:2022myb}.

Beyond the leading twist, power-suppressed contributions $\sim (\sqrt{-t}/Q)^k $ and $\sim (m/Q)^k$, 
where $t$ is the invariant momentum transfer and $m$ is the target mass, play a special role. They can be large and have to be taken into account.
Indeed, the transverse spatial position of partons in the target is Fourier conjugate to the momentum transfer in the scattering process. 
Hence the resolving power of DVCS is directly limited by the range of the invariant moment transfer $t$ which can be used in the analysis. 
Theoretical control over power corrections $(\sqrt{-t}/Q)^k$ is therefore crucial for three-dimensional imaging. 
One more pressing issue is to clarify whether target mass corrections  $\sim (m/Q)^k$
do not endanger QCD factorization for coherent DVCS on nuclei \cite{CLAS:2017udk,CLAS:2021ovm}.

We refer to the $\sim (\sqrt{-t}/Q)^k $ and $\sim (m/Q)^k$ contributions to Compton amplitudes as ``kinematic power corrections'' because
they do not involve new nonperturbative inputs in addition to the leading twist GPDs. On an intuitive level, their origin and
interpretation can be explained as follows~\cite{Braun:2014paa}.
The four-momenta of the initial and final photons and nucleons in a DVCS process do not  
lie in one plane. Hence the distinction of longitudinal and transverse directions is convention-dependent and, as a consequence,
the leading-twist approximation is intrinsically ambiguous. In the Bjorken limit this is a  $1/Q$ effect. 
On a more technical level, the freedom to redefine large ``plus'' parton momenta by adding small transverse components has two
consequences. First, the dependence  of the skewness parameter $\xi$ on the Bjorken variable $x_B$ acquires $t$-dependent 
power suppressed contributions. Second, such a redefinition generally leads to excitations of the subleading photon helicity-flip
amplitudes~\cite{Braun:2014sta,Braun:2014paa}. This convention-dependence proves to be 
rather large, see~\cite{Guo:2021gru} for a detailed study. It should be viewed as a theoretical uncertainty that can and should be 
removed by explicit calculation of the kinematic power corrections and adding them to the leading-twist expressions in the data analysis. 

This problem was addressed in~\cite{Braun:2011zr,Braun:2011dg,Braun:2012bg,Braun:2012hq}, where a technique was developed that allows one to 
calculate kinematic corrections to the twist-four accuracy, i.e. up to terms $\sim t/Q^2$ and $\sim m^2/Q^2$. 
The results in the final form were presented in~\cite{Braun:2014sta}.  
A typical size of kinematic corrections for $|t|/Q^2 \lesssim 1/4$ was found to be of the order of 10\% for asymmetries, 
but they could be as large as 100\% for the DVCS cross section in certain kinematics. 
These corrections can significantly impact the extraction of GPDs from  data and have to be taken into
account~\cite{Defurne:2015kxq,Defurne:2017paw,JeffersonLabHallA:2022pnx}.

\allowdisplaybreaks

The approach of~\cite{Braun:2011zr,Braun:2011dg,Braun:2012bg,Braun:2012hq} 
requires explicit construction of the higher-twist operator basis and becomes unwieldy beyond twist four.
In Ref.~\cite{Braun:2020zjm} we suggested a different technique to calculate kinematic corrections to generic two-photon processes, based on the conformal field theory (CFT) methods.
This technique is  more general and is applicable to all twists. Using this new approach we have calculated in Ref.~\cite{Braun:2022qly} the kinematic power corrections
to twist-six accuracy for the simplest case of DVCS on a scalar target. 
In this work we 
derive the corresponding expressions for the spin-1/2 targets (nucleon). We achieve the following accuracy, schematically:
\begin{align}
       \mathcal{A}^{(\pm,\pm)} &\sim 1 + \frac{1}{Q^2} + \frac{1}{Q^4}\,,
\notag\\
       \mathcal{A}^{(\pm,0)} &\sim \frac{1}{Q} + \frac{1}{Q^3} \,,
\notag\\
       \mathcal{A}^{(\pm,\mp)} &\sim \frac{1}{Q^2} + \frac{1}{Q^4} \,,
\label{intro:2}
\end{align}
where $\mathcal{A}^{(\pm,\pm)}$, $\mathcal{A}^{(\pm,0)}$ and $\mathcal{A}^{(\pm,\mp)}$ are the helicity-conserving, helicity-flip and double-helicity-flip
amplitudes, respectively. Precise definitions are given in the text.
Taking into account these corrections removes the frame dependence of the leading-twist
approximation and restores the electromagnetic gauge invariance of the Compton amplitude up to $1/Q^5$ effects.

Besides providing general expressions, 
we study the numerical impact of kinematic corrections on several key experimental DVCS observables. 
We find that the twist-five and twist-six contributions can be decreased by changing the expansion parameter 
in the twist-three and twist-four corrections from the photon virtuality $Q^2$ to $Q^2+t$. In addition, we confirm the 
observation made in~\cite{Braun:2014sta} that target mass corrections always involve powers of the skewness parameter $\sim(\xi m/Q)^k$      
and do not endanger QCD factorization for coherent DVCS on nuclei.


                                   \section{BMP helicity amplitudes}



                                   \subsection{Definitions and conventions}\label{sec:BMP}


In this work we consider DVCS on the nucleon target
\begin{align}
\gamma^*(q)+ N(p,s) \longrightarrow \gamma(q')+ N(p',s')\,.
\label{DVCSprocess-BMP}
\end{align}
The DVCS amplitude $\mathcal{A}_{\mu\nu}$ is defined as:
\begin{align}
\label{Amunu-def}
 &\mathcal{A}_{\mu\nu}(q,q',p)=\\
 &=i\! \int\!\! d^4 x\, e^{-i(z_1q-z_2 q')\cdot x }
\langle p',s'|T\{j_\mu(z_1x)j_\nu(z_2x)\}|p,s\rangle,
\nonumber
\end{align}
where $j_\mu(z_1x)$ and $j_\nu(z_2x)$ are the electromagnetic currents, $z_1,z_2$ are real numbers such that
$z_1-z_2=1$. Note that $\mathcal{A}_{\mu\nu}$ does not depend on $z_1+z_2$.
This property is referred to as translation invariance in Refs.~\cite{Braun:2012hq,Braun:2012bg}. 
It is violated at the leading twist and is restored by adding kinematic power corrections to the required accuracy.

We follow the BMP convention~\cite{Braun:2012hq,Braun:2012bg} and use the photon momenta, $q$ and  $q'$, to define a 
longitudinal plane spanned by the two light-like vectors
\begin{align}
n=q'\,, \qquad \tilde n=-q+(1-\tau)\, q'\,,
\end{align}
where $\tau= t/(Q^2+t)$ with $Q^2=-q^2$.
For this choice the momentum transfer to the target
$$\Delta = p'-p= q-q'\,, \qquad t=\Delta^2$$
is purely longitudinal and both --- initial and final state --- proton momenta have a nonzero
transverse component
\begin{align}\label{Pperp}
P_\mu&=\frac12\left(p+p^\prime\right)=\frac{1}{2\xi}\left(\bar n_\mu-\tau n_\mu\right) + P_{\perp,\mu}\,.
\end{align}
The skewness parameter $\xi$ is defined as
\begin{align}
\xi \equiv \xi^{\rm BMP} &= -\frac{\Delta\cdot q'}{2P\cdot q'}
= \frac{x_B(1+t/Q^2)}{2-x_B(1-t/Q^2)}
\label{xiBMP}
\end{align}
and $|P_\perp|^2$ can  be written in terms of kinematic invariants as
\begin{align}
 |P_\perp |^2 = \frac{1-\xi^2}{4\xi^2} (t_{\rm min}-t)\,, && t_{\rm min} = -\frac{4m^2\xi^2}{1-\xi^2} \,.
\end{align}
The BMP choice is advantageous mainly because it leads to simple expressions for the photon polarization vectors that can be  chosen as follows:
\begin{align}
\varepsilon^0_\mu&=-\left(q_\mu-q'_\mu {q^2}/{(q\cdot q')}\right)/{\sqrt{-q^2}}\,,\notag\\
\varepsilon^\pm_\mu&=(P^\perp_\mu\pm i \bar P^\perp_\mu)/ {(\sqrt{2}|P_\perp|)}\,,
\label{BMPpolarizations}
\end{align}
where $P^\perp_\mu=g_{\mu\nu}^\perp P^\nu$, $\bar P^\perp_\mu=\epsilon_{\mu\nu}^\perp P^\nu$ and
\begin{align}
g_{\mu\nu}^\perp&=g_{\mu\nu}-(q_\mu q'_\nu+q'_\mu q_\nu)/(q\cdot q')+{q'_\mu}q'_\nu\,{q^2}/(q\cdot q')^2\,,
\notag\\
\epsilon_{\mu\nu}^\perp&=\epsilon_{\mu\nu\alpha\beta}{q^\alpha q'^\beta}/(q\cdot q')\,,\qquad {\epsilon^{\rm BMP}_{0123}=1}\,.
\end{align}
Normalization is such that $\varepsilon^+_\mu\varepsilon^{-\mu} = -1$\,,
$\varepsilon^0_\mu\varepsilon^{0\mu} = +1$. The pair $\varepsilon^\pm_\mu$ form 
a basis 
in the transverse plane whereas $\varepsilon^0_\mu$
is a unit vector in longitudinal plane 
orthogonal to the 
photon momentum
$q'$.

The DVCS amplitude $\mathcal{A}_{\mu\nu}$ can be decomposed in terms of scalar (helicity) amplitudes using this basis:
\begin{align}
\!\mathcal{A}_{\mu\nu}={} &\varepsilon^+_{\mu} \varepsilon^-_{\nu} \mathcal{A}^{++}
+\varepsilon^-_{\mu} \varepsilon^+_{\nu} \mathcal{A}^{--}
+\varepsilon^0_{\mu} \varepsilon^-_{\nu} \mathcal{A}^{0+}
\notag\\
{}+{}&\varepsilon^0_{\mu} \varepsilon^+_{\nu} \mathcal{A}^{0-}\!
+\!\varepsilon^+_{\mu} \varepsilon^+_{\nu} \mathcal{A}^{+-}\!
+\!\varepsilon^-_{\mu} \varepsilon^-_{\nu} \mathcal{A}^{-+} 
\!.
\label{Amunu}
\end{align}
We neglected a term proportional to $q'_\nu $ since it does not contribute to any
observable. Each helicity amplitude involves the sum over quark flavors, $\mathcal{A}=\sum e_q^2
\mathcal{A}_q$, and is written in terms
of the leading-twist GPDs $H^q,E^q,\widetilde H^q,\widetilde E^q$. For the GPD definitions
(see below), we follow Ref.~\cite{Diehl:2003ny}.


                                \subsection{Light-ray OPE}


The amplitude 
$\mathcal{A}_{\mu\nu}$ can be written in terms of matrix elements of 
$C=+1$ 
twist-2 quark-antiquark light-ray operators 
\begin{flalign}\label{OVA}
\mathscr O_V(z_1y,z_2y) &=\frac12\langle p'| [\bar q(z_1 y)\slashed y  q(z_2 y)]_{\ell t} \! - (z_1\!\leftrightarrow\! z_2)
|p\rangle, 
\notag\\
\mathscr O_A(z_1y,z_2y) &=\!\frac12\langle p'|[\bar q(z_1 y)\slashed y \gamma_5  q(z_2 y)]_{\ell t }\! +\!
(z_1\!\leftrightarrow\! z_2)|p\rangle,
\end{flalign}
where on  the l.h.s. only the dependence on the quark positions is shown in order not to overload 
notation.
In these expressions the Wilson line connecting the quarks is implied,
 and the notation $[\ldots]_{\ell t}$ stands for the 
leading twist projection as defined in Ref.~\cite{Balitsky:1987bk}.
The matrix elements \eqref{OVA} can be written in terms of the GPDs as follows:
\begin{align}\label{defGPD}
\mathscr{O}_V(z_1y,z_2y)&=
\int_{-1}^1\! dx  \big[y^\rho e^{-i (Py)[z_1(\xi-x)+z_2(x+\xi)]}\big]_{\ell.t.}
\notag\\&\quad \times
\Big\{ h_\rho   H(x,\xi,t) + e_\rho E(x,\xi,t)\Big\},  
\notag\\
\mathscr{O}_A(z_1y,z_2y) &=
\int_{-1}^1\!\! dx\, \big[y^\rho e^{-i (Py)[z_1(\xi-x)+z_2(x+\xi)]}\big]_{\ell.t.}
\notag\\&\quad\times
\Big\{\tilde h_\rho  \widetilde H(x,\xi,t) + \tilde e_\rho\widetilde E(x,\xi,t) \Big\}.  
\end{align}
In these expressions we use 
short-hand notations \cite{Belitsky:2012ch} for the Dirac spinor bilinears
\begin{align}
& h_\rho = \bar u(p') \gamma_\rho u(p)\,, 
&& \tilde h_\rho = \bar u(p') \gamma_\rho\gamma_5 u(p)\,, 
\notag\\
&e_\rho = \bar u(p')\frac{i\sigma^{\rho\alpha}\Delta_\alpha}{2m} u(p)\,,
&& \tilde e_\rho = \frac{\Delta_\rho}{2m} \bar u(p') \gamma_5 u(p)\,. 
\label{Dirac}
\end{align}
The leading twist projection of the exponential function is given by~\cite{Balitsky:1990ck}
\begin{align}
[e^{-i\ell y}]_{lt} &= e^{-i\ell xy} + \frac14 y^2 \ell^2  \int_0^1\!dt\,t\,   e^{-it\ell y}
\notag\\&\quad
+ \frac{1}{32} y^4 \ell^4 \int_0^1\!dt\,\bar t\, t^2  e^{-it\ell y} +\mathcal{O}(y^6)\,,
\notag\\
[y^\rho e^{-i\ell y}]_{lt} &= i \frac{\partial}{\partial \ell_\rho} [e^{-i\ell y}]_{lt}\,,
\qquad
\bar t = 1-t\,. 
\end{align}

The calculation of the DVCS amplitude in terms of the matrix elements
of light-ray operators uses conformal symmetry techniques and
is explained in Refs.~\cite{Braun:2020zjm,Braun:2022qly}. 
Separating the contributions of the vector and axial-vector operators, $\mathcal A^{\mu\nu}=\mathcal A^{\mu\nu}_V +\mathcal A^{\mu\nu}_A$, we obtain
\begin{widetext}
\begin{align}\label{fin-vector}
\mathcal A^{\mu\nu}_V & =
\int \frac{d^4x}{\pi^2} \frac{e^{-iqx}}{(-x^2+i0)}\int_0^1d\alpha\int_0^{\bar\alpha}d\beta
\biggl\{\frac{1}{(-x^2+i0)}
\Big(g^{\mu\nu}\delta(\alpha)\delta(\beta)-x^\mu\partial^\nu \delta(\beta) - x^\nu \nabla^\mu \delta(\alpha)\Big)\mathscr O_V
 -\frac{i}{2} 
 \big( \Delta^\nu\partial^\mu  -\Delta^\mu \partial^\nu\big) \mathscr O_V
\notag\\&\quad
+\frac14 g^{\mu\nu} \Big( \mathscr O_V^{(1)} - \delta(\alpha)\mathscr O_V^{(2)}\Big)
 +\frac14 \big(x^\nu\partial^\mu  {+}x^\mu\nabla^\nu\big)
 \Big(\ln\bar\tau\,\mathscr O_V^{(1)} +\frac{\beta}{\bar \beta}\,\mathscr O_V^{(2)}\Big)
+\frac12 
\big(x^\nu\partial^\mu   -  x^\mu \nabla^\nu \big)
 \frac\tau{\bar\tau}
\Big(-\mathscr O_V^{(1)} +\frac{\bar\alpha}{\alpha}\,\mathscr O_V^{(2)}\Big)
\notag\\&\quad
-\frac14 x^\nu\nabla^\mu
 \frac\beta{\bar\beta}\biggl[{4}\Big(\frac12+\frac{\tau}{\bar\tau}\Big)\,\mathscr O_V^{(1)}
-\Big(\delta(\alpha)+\frac{\beta}{\bar\beta}\Big)\mathscr O_V^{(2)}\biggr]
-\frac{x^\mu x^\nu}{(-x^2+i0)} 
\biggl[
\big(\ln\bar\tau+\ln\bar\alpha +1 \big)\,\mathscr O_V^{(1)} + \frac{\beta}{\bar \beta}\,\mathscr O_V^{(2)}
\biggr]
\notag\\&\quad
+ \frac12 x^\mu\partial^\nu 
\biggl[
\ln\bar\alpha\,\mathscr O_V^{(1)} +  \Big(\frac12-\frac{2\tau}{\bar\tau}\Big) \mathscr O_V^{(2)}
\biggr]
+\frac{x^\mu x^\nu}4
\biggl[
\Big(i(\Delta\partial) +\frac{\Delta^2}2\Big) \frac{\beta}{\bar\beta}\Big( \frac2{\bar\tau}-1\Big)
-2\Big(i(\Delta\partial) +\frac{\Delta^2}4\Big)
\Big(\ln\bar\tau+\frac{2\tau}{\bar\tau}\Big)
\biggr] \mathscr O_V^{(1)}
\biggr\},
\intertext{and}
\label{fin-axial}
\mathcal A_A^{\mu\nu} &=
\frac{1}{2}\int \frac{d^4x}{\pi^2}\frac{ e^{-iqx}}{(-x^2+i0)}\int_0^1d\alpha\int_0^{\bar\alpha} d\beta
\biggl\{
{ i\epsilon_{\mu\nu\beta\gamma} x^\beta}
\biggl[
\frac1{(-x^2+i0)} \Big(-\nabla^\gamma\delta(\alpha) -\partial^\gamma \delta(\beta) \Big)\mathscr O_A
\notag\\&\quad
+\frac14\nabla^\gamma\biggl( \Big(\ln\bar\tau- \frac{2\beta}{\bar\beta}\Big) \mathscr O_A^{(1)}
+\frac{\beta}{\bar\beta}\Big(1+\delta(\alpha)-\frac{\beta}{\bar\beta}\Big)\mathscr O_A^{(2)}
\biggr)
 +\frac14 \partial^\gamma  
 \biggl(
 \big(\ln\bar\tau+ {2}\ln\bar\alpha \big)\,\mathscr O_A^{(1)} 
+
\frac{1}{\bar \beta} \mathscr O_A^{(2)}
\biggr)
\biggr]
\notag\\&\quad
+ \big(x_{\nu}\epsilon_{\mu\alpha\beta\gamma} +x_{\mu} \epsilon_{\nu\alpha\beta\gamma}\big)
x^\alpha \Delta^\gamma \partial^\beta
\biggl[\frac1{(-x^2+i0)}
\mathscr O_A
-\frac14 
\Big(\ln\bar\tau\,\mathscr O_A^{(1)}
+\frac{\beta}{\bar \beta}\,\mathscr O_A^{(2)}
\Big)
\biggr]
\biggr\}\,,
\end{align}
\end{widetext}
where
\begin{align}
\mathscr O_{X}^{(1)}(z_1x,z_2x) &= i(\Delta\partial) \mathscr O_{X}(z_1x,z_2x)\,,
\notag\\
\mathscr O_{X}^{(2)}(z_1x,z_2x) &=\left(i(\Delta\partial)+ \frac{\Delta^2}2\right) \mathscr O_{X}(z_1x,z_2x)\,,
\end{align}
with $X=A,V$
and
\begin{align}
 \tau = \frac{\alpha\beta}{\bar\alpha\bar\beta}\,,
\qquad \partial^\gamma = \frac{\partial}{\partial x_\gamma}\,, 
\qquad  \nabla^\gamma = \partial^\gamma- i\Delta^\gamma.
\end{align}
 For brevity, we do not show the arguments of the operator matrix elements, which are the same
for all cases, $\mathscr O_X$ stands for $\mathscr O_X (\bar\alpha x,\beta x)$. 
The expression in~\eqref{fin-vector} is equivalent to that given in Ref.~\cite{Braun:2022qly}; 
the contribution of axial-vector operators in \eqref{fin-axial} is a new result.


                                                  \subsection{Results}\label{sec:results}


In this section we denote vector and axial-vector Dirac bispinors as
\begin{align}\label{Dirac:va}
v^\mu&=\bar u(p')\gamma^\mu u(p) \,, 
\notag\\
a^\mu&=\bar u(p')\gamma^\mu \gamma_5u(p)\,,
\end{align}
and use shorthand notations
\footnote{In Ref.~\cite{Braun:2012hq} there is a sign error in the in-line equation after Eq.(7), it is corrected in~\cite[Eq.(A11)]{Braun:2014sta}.} 
for the scalar products with the BMP polarization vectors
defined in \eqref{BMPpolarizations}:
\begin{align}\label{Dirac2}
 v_\perp^\pm&=(v\cdot\varepsilon^\pm)\,,\quad a_\perp^\pm=(a\cdot\varepsilon^\pm)\,,
\notag\\
P_\perp^\pm&=(P\cdot \varepsilon^\pm) = - |P_\perp|/\sqrt{2}\,.
\end{align}
%


At the intermediate stages of the calculation the ``double distribution'' (DD) \cite{Radyushkin:1997ki} parametrization 
of the nucleon matrix elements of light-ray vector- and axial-vector operators proves to be the most convenient.
The results in the DD representation are collected in appendix~\ref{app:DD}.

In the following expressions we use rescaled variables
\begin{align}
  \t = \frac{t}{(qq')}\,, 
\qquad \mm = \frac{m^2}{(qq')}\,, 
\qquad \Pperp = \frac{| P_\perp |^2}{(qq')},
\label{rescale}
\end{align}
where $(qq') = - 1/2 (Q^2+t)$. We use the notation
\begin{align}
M(x,\xi,t) = H(x,\xi,t) + E(x,\xi,t)\,,
\end{align}
and 
\begin{align}
D_\xi =(-2\xi^2\partial_\xi)\,.
\end{align}
It is convenient to use different normalization for the convolution integrals
of the coefficient functions with different GPDs:
\begin{align}
(M \otimes T) & = \int dx\, M(x,\xi,t)\, T\Big(\frac{x+\xi}{2\xi}\Big), 
\notag\\
(\widetilde H \otimes T)  &= \int dx\, \widetilde H (x,\xi,t)\, T\Big(\frac{x+\xi}{2\xi}\Big),
\notag\\
(E \odot T) & = \frac1{2\xi} \int dx\, E(x,\xi,t)\, T\Big(\frac{x+\xi}{2\xi}\Big), 
\notag\\
(\widetilde E \circledast T)  &= \frac12 \int dx\, \widetilde E (x,\xi,t)\, T\Big(\frac{x+\xi}{2\xi}\Big).
\label{convolutions}
\end{align}
The coefficient functions appearing in the equations below are defined as
\begin{align}
T_0(z) &=\frac1{\bar z}\,,
\hspace*{2cm}
T_{1}(z)= \ln\bar z\,,
\notag\\
T_{00}(z)& =\frac{\bar z} z \ln \bar z\,,
\qquad\qquad
T_{10}(z)= \frac1 z \ln\bar z\,,
\notag\\
T_{11}(z)&=\Big( 2-\frac1z\Big)\ln\bar z\,,
\notag\\
 T_V(z)&=
\frac1{\bar z}\big({\Li_2(z)-\zeta_2}\big)-\ln\bar z\,,
\notag\\
 T_A(z)&=-\frac1{\bar z}\big({\Li_2(z)-\zeta_2}\big) + \frac1z\ln\bar z\,,
\notag\\
 T_2(z)&=\frac1{\bar z}\big(\Li_2(z)-\zeta_2\big) -\frac1{2z}\ln\bar z\,,
\notag\\
 T_3(z)&=\frac{2z+1}{\bar z}\big(\Li_2(z)-\zeta_2\big) -\frac1{2}\Big(7-\frac1z\Big)\ln\bar z\,.
\label{Tfunctions}
\end{align}
They are analytic functions of $z$ with a cut from 1 to $\infty$, apart from $T_0$ which has a pole singularity. 
Functions of higher transcendentality appear at intermediate steps of the calculation but cancel in the final expressions. 
The convolution integrals \eqref{convolutions} involve the CFs on the upper side of the cut: $T(z) \mapsto T(z+i\epsilon)$ for $x >\xi>0$. 
%
\subsubsection{Helicity-conserving $(\pm,\pm)$ amplitude}
We obtain
\begin{align}
\mathcal A_V^{\pm,\pm}  &=\frac{ (vq') }{(qq')} V^{(1)}_0 +\frac{(vP)}{m^2} V^{(2)}_0\,, 
\notag\\
\mathcal A_A^{\pm,\pm}  &= \pm\frac{(aq')}{(qq')}  A^{(1)}_0 \pm \frac{(a\Delta)}{2m^2}  A^{(2)}_0\,,
\label{Apmpm}
\end{align}
where
\begin{widetext}
\allowdisplaybreaks{
\begin{align}
 V^{(1)}_0 &=
-\left(1+\frac{\t}{4}\right)  \Big(M\otimes T_0\Big) -\frac \t 2 \Big(M\otimes T_{10}\Big)
  +\frac14 \t^2 \Big(M\otimes T_{11}\Big) -\frac12 D_\xi^2 \Pperp
 \Big(M\otimes\big( T_2 + 2 \t\, T_V\big)\Big)
+\frac18 D_\xi^3 |\widehat{P}_\perp|^4 D_\xi\Big(M\otimes T_3\Big)\,,
\notag\\
 V^{(2)}_0 &=
-\left(1+\frac{\t}{4}\right)  \Big(E\odot T_0\Big) -\frac{\t}{2}\Big(E\odot T_{10}\Big) +\frac{\t^2}{4} \Big( E\odot T_{11}\Big)
-\frac12 D_\xi \Pperp
D_\xi \Big(E\odot \big(T_2+2\t\, T_V\big)\Big)
+ \frac18 D_\xi^2 |\widehat{P}_\perp|^4 D_\xi^2 \Big(E\odot T_3\Big)
\notag\\&\quad
-\mm\biggl\{ { D_\xi \Big( M \otimes \big( T_2 + 2\t T_V\big)\Big)}
-\frac12 D_\xi^2 \Pperp
 \Big(M\otimes T_3\Big)
\biggr\}\,,
\notag\\
 A^{(1)}_0&
=\left(1+\frac{\t}{4}\right)  \Big(\widetilde H \otimes T_0\Big) + \frac \t 2\left(1+\frac \t 2 \right) \Big(\widetilde H \otimes T_{10}\Big)
+\frac12 D_\xi^2\Pperp \Big(\widetilde H\otimes \big(T_2-2\t\, T_A\big)\Big)
-\frac3{16}D_\xi^3 |\widehat{P}_\perp|^4 D_\xi\left( \widetilde H\otimes \big(T_{00} -2  T_A\big)\right)\,,
\notag\\
 A^{(2)}_0 &=\left(1+\frac{\t}{4}\right)  \Big(\widetilde E\circledast T_0\Big) + \frac{\t}{2}  \Big(\widetilde E\circledast T_{10}\Big)
+\frac{{\widehat t}^2}4 \Big(\widetilde  E\circledast \Big(T_{10} + \frac32\Big)\Big)
+\frac12 D_\xi \Pperp D_\xi\Big(\widetilde E\circledast \big(T_2- 2\t\, T_A\big)\Big)
\notag\\&\quad
-\frac 3{16} D_\xi^2 |\widehat{P}_\perp|^4 D_\xi^2 \Big(\widetilde E\circledast \big(T_{00}-2T_A\big)\Big)
+\mm D_\xi  \biggl\{\frac1\xi
\Big(\widetilde H \otimes \big(T_2-2\t T_A\big)\Big)
-\frac34 D_\xi\frac1\xi \Pperp D_\xi\Big(\widetilde H \otimes \big(T_{00}-2T_A\big)
\Big)
\biggr\}.
\end{align}
}
\end{widetext}
 We have verified that the invariant amplitudes $(V_0^{(1)},\ V_0^{(2)}) $ and $(A^{(1)}_0,\ A^{(2)}_0)$ coincide with  $(\mathbb V_{2}/2,\ \mathbb V_1)$ and
 $(\mathbb A_{2}/2,\  \mathbb A_1)$ as defined in Ref.~\cite[Eq.(A15)]{Braun:2014sta}, up to twist-six terms $\t^2, \t\,\mm,\ldots$, respectively.

\subsubsection{Helicity-flip $(0,\pm)$ amplitude}
%
This is a subleading-power amplitude that starts at the twist-3 level. We obtain
\begin{align}
\mathcal A_V^{0,\pm}  &=\frac{Q}{(qq')}
\left({v_\pm }\, V_1^{(1)} +\frac{(vq')P_\pm}{(qq')}\, V_1^{(2)} + \frac{(vP)P_\pm}{m^2}\, V_1^{(3)}\right),
\notag\\
\mathcal A_A^{0,\pm}  &=\pm\frac{Q}{(qq')}\left({a_\pm}\, A_1^{(1)} +
 \frac{(aq')P_\pm}{(qq')}\, A_1^{(2)} + \frac{(a\Delta)P_\pm}{2m^2}\, A_1^{(3)}\right)
\label{A0pm}
\end{align}
with
\begin{align}
 V_1^{(1)} &=-\left(1+\frac \t 2\right)\left(M\otimes T_{10}\right) + \t \,\left(M\otimes T_{1}\right)
\notag\\&\quad
  -\frac12 D_\xi^2 \Pperp \left(M\otimes T_V\right),
\notag\\
 V_1^{(2)}&=- D_\xi V_1^{(1)},
\notag\\
 V_1^{(3)} &=\left(1+\frac \t 2\right)D_\xi\left( E\odot T_{10}\right) - \t\, D_\xi\left( E\odot T_{1}\right)
\notag\\ &\quad
      +\frac12 D_\xi^2 \Pperp D_\xi\Big( E\odot T_V\Big)  +\mm D_\xi^2\Big(M\otimes T_V\Big),
\notag\\
 A_1^{(1)} &=\left(1+\frac \t 2\right)\left( \widetilde H\otimes T_{10}\right) 
 -\frac12 D_\xi^2\Pperp\Big( \widetilde H\otimes T_A\Big)\,,
\notag\\
A_1^{(2)}& =-D_\xi A_1^{(1)},
\notag\\
 A_1^{(3)} &=-\left(1+\frac \t 2\right)D_\xi\left(\widetilde E\circledast T_{10}\right)
+\frac12 D_\xi^2 \Pperp D_\xi \Big(\widetilde  E\circledast T_A\Big)
\notag\\
&\quad
+\mm D_\xi^2 \frac1\xi \Big( \widetilde H\otimes T_A\Big).
\end{align}
To twist-three accuracy (leading terms), these expressions agree with Ref.~\cite[Eq.(A16)]{Braun:2014sta}.

\subsubsection{Double-helicity-flip $(\mp,\pm)$ amplitude}
%
\begin{align}
\mathcal A_V^{\mp\pm}  &=\frac{v_\pm P_\pm}{(qq')}\, V_2^{(1)} +\frac{(vq')}{(qq')}\, V_2^{(2)} + \frac{(vP)}{m^2}\, V_2^{(3)},
\notag\\
\mathcal A_A^{\mp\pm}  &=\pm\frac{a_\pm P_\pm}{(qq')}\, A_2^{(1)} \pm \frac{(aq')}{(qq')}\, A_2^{(2)} \pm \frac{(a\Delta)}{2m^2}\, A_2^{(3)}.
\label{Amppm}
\end{align}
We get 
\begin{align}
V_2^{(1)} &=2\left(1+\frac{\t}{4}\right) D_\xi \, \Bigl(M\otimes T_{11}\Bigr)  
-\frac12 D_\xi^3 \Pperp  \Big(M \otimes  T_V\Big)\,,
\notag\\
V_2^{(2)}&=-\frac14 \Pperp D_\xi V_2^{(1)},
\notag\\
V_2^{(3)} &=-\frac12\Pperp D_\xi^2\biggl\{\left(1+\frac{\t}{4 }\right)   \, \Bigl(E\odot T_{11}\Bigr)
 \notag\\
 &\quad       -\frac14 D_\xi \Pperp D_\xi
           \Bigl(E\odot  T_{V}\Bigr) 
-\frac12 \mm \, D_\xi \Bigl(M \otimes  T_V\Bigl)\biggr\},
\notag\\
A_2^{(1)} &=2\left(1+\frac{\t}{4}\right) D_\xi \, \Bigl(\widetilde H\otimes T_{10}\Bigr) 
-\frac12 D_\xi^3\Pperp \Big(\widetilde H \otimes  T_A\Big)\,,
\notag\\
A_2^{(2)}& =-\frac14 \Pperp D_\xi A_2^{(1)},
\notag\\
A_2^{(3)} &=-\frac12\Pperp D_\xi^2\biggl\{\left(1+\frac{\t}{4 }\right)   \, \Bigl(\widetilde E\circledast T_{10}\Bigr)
\notag\\
&\quad  -\frac14 D_\xi \Pperp D_\xi 
\Bigl(\widetilde E\circledast  T_{A}\Bigr) -\frac12 \mm D_\xi \frac1\xi \Bigl(\widetilde H \otimes  T_A\Bigl)\biggr\}.
\end{align}
At leading order (twist four), these expressions agree with the corresponding results in~\cite[Eq.(A17)]{Braun:2014sta}, 
except for the sign of the $ A_2^{(3)}$ contribution. In the DD representation~\cite[Eq.(B11)]{Braun:2014sta} all results agree.

One of the motivations for our study was to clarify whether target mass corrections  $\sim (m/Q)^k$
do not endanger QCD factorization for coherent DVCS on nuclei \cite{CLAS:2017udk,CLAS:2021ovm}.
By inspection of the above equations, one can check that target mass dependent contributions always involve
additional factors of the skewness parameter, so that the expansion goes in powers 
of $\xi^2 m^2/Q^2$ rather than $m^2/Q^2$. For nuclear targets, effectively, $m\to A m$ and $\xi \to \xi/A$, 
so that the target mass corrections remain essentially the same as for the nucleon and are small, 
apart from the large $x_B$ region.


                            \section{Compton form factors and the frame dependence}

Compton form factors (CFFs) are defined \cite{Belitsky:2001ns} through the decomposition of the helicity amplitudes in terms of the set of bilinear spinors
in Eq.~\eqref{Dirac}
\begin{align}
\mathcal A^{a,\pm}_{\BMP} &
= \mathcal H^{a,\pm}_{\BMP} h + \mathcal E^{a,\pm}_{\BMP} e 
\mp \widetilde{\mathcal H}^{a,\pm}_{\BMP} \tilde h 
\mp \widetilde{\mathcal E}^{a,\pm}_{\BMP} \tilde e\,,   
\end{align} 
where $a = (+,0,-)$ and we have reintroduced the notation ``BMP'' to remind that the the helicity amplitudes and 
hence also the CFFs are defined using BMP conventions, see section~\ref{sec:BMP}.
Making use of the Dirac equation for the nucleon states, one finds \cite{Braun:2014sta}
\begin{align}
   \frac{(vP)}{2m^2} &= h - e\,,
&&
   \frac{(vq')}{qq'} = -\frac{1}{\xi} h\,,
\notag\\
   \frac{(a\Delta)}{4m^2} &= -\frac{1}{\xi}\left(\! 1+ \frac{t}{Q^2}\!\right)\tilde e\,,
&&
   \frac{(aq')}{(q q')} = - \frac1{\xi}\tilde h -\frac{1}{\xi} \frac{4m^2}{Q^2}\tilde e \,,
\label{BMJ2BMPspinors1}
\end{align}
and
\begin{align}
  \frac{v_\perp^\pm}{\sqrt{2}}&=
 - |P_\perp|h - \frac{m^2}{|P_\perp|}\biggl[e - \frac{t}{4m^2} h\biggr]
 \mp \frac{m^2}{\xi|P_\perp|}\biggl[\tilde e - \frac{t}{4m^2}\, \tilde h\biggr],
\notag\\
 \frac{a_\perp^\pm}{\sqrt{2}} &=
  -\frac{m^2}{\xi^2 |P_\perp|}\biggl[\tilde e - \frac{t}{4m^2} \tilde h\biggr]
\mp \frac{m^2}{\xi |P_\perp|}\biggl[e - \frac{t}{4m^2} h\biggr],
\label{BMJ2BMPspinors2}
\end{align}
where $\xi \equiv \xi_{\BMP}$ is defined in Eq.~\eqref{xiBMP}.
Making use of these relations it is straightforward to obtain the expressions 
for the CFFs 
as linear combinations of the invariant functions 
$V_{0,1,2}^{(1,2,3)}$, $A_{0,1,2}^{(1,2,3)}$ from the previous section.    

In the DVCS phenomenology a different decomposition of the Compton amplitude is traditionally used, 
being a certain generalization of the standard DIS reference
frame where the initial photon and proton momenta form the
longitudinal plane. Several conventions based on this identification exist and are in practice 
very similar to each other. The KM convention used by Kumericki and M{\"u}ller in global DVCS leading-twist fits
is one such example.
Belitsky, M{\"u}ller and Ji (BMJ) \cite{Belitsky:2012ch} used the KM decomposition to derive explicit 
expressions for key DVCS observables including subleading-power CFFs.

The main difference to BMP conventions is that KM (and BMJ)  define helicity amplitudes in the target rest frame and 
to this end, introduce different sets of polarization vectors for the initial and final state photons.
Also the definition of the skewness variable is different, $\xi_{\BMJ}= x_B/(2-x_B)$.
The relation between the BMP and KM (BMJ) CFFs is just a Lorentz transformation and can easily be worked out,
see \cite{Braun:2014sta}:
\begin{align}
\label{BMJ-BMP}
{\mathcal F}^{\pm+}_{\BMJ} &= {\mathcal F}^{\pm+}_{\BMP} + \frac{\varkappa}{2}\Big[{\mathcal F}^{++}_{\BMP} + {\mathcal F}^{-+}_{\BMP}\Big]- \varkappa_0\,
 {\mathcal F}^{0+}_{\BMP},
\\
 {\mathcal F}^{0+}_{\BMJ} &=  - \left(1+\varkappa\right)  {\mathcal F}^{0+}_{\BMP}  
+ \varkappa_0\Big[{\mathcal F}^{++}_{\BMP} + {\mathcal F}^{-+}_{\BMP}\Big],
\notag
\end{align}
etc.
Here all entries $\mathcal F \in \{\mathcal H, \mathcal E,\widetilde{\mathcal H},\widetilde{\mathcal E}\}$
are functions of $x_B,t,Q^2$. Also  
\begin{align}
\varkappa_0 &= \frac{\sqrt{2} Q \widetilde K}{\sqrt{1 + \gamma^2}(Q^2+t)}  = \mathcal{O}(1/Q)\,,
\notag\\
\varkappa &=\frac{{ Q}^2  - t + 2 x_B t}{\sqrt{1 + \gamma^2}(Q^2+t)}-1   = \mathcal{O}(1/Q^2)\,,
\label{varkappa}
\end{align}
where 
\begin{align}
\gamma = 2m x_B/Q\,, \qquad  \widetilde K &= x_B(1+t/Q^2) |P_\perp|\,.
\end{align}
%
These relations are exact, there is no approximation. 

\begin{figure*}[t]
\includegraphics[width=0.31\textwidth, clip = true]{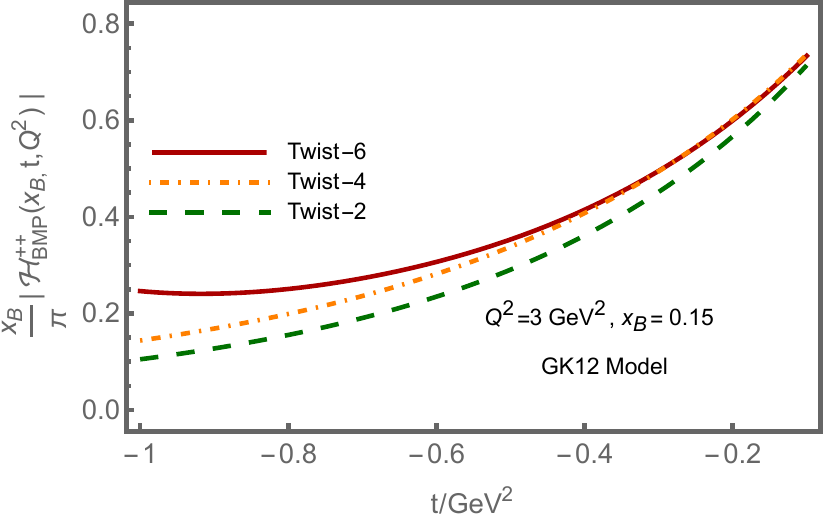}
\includegraphics[width=0.31\textwidth, clip = true]{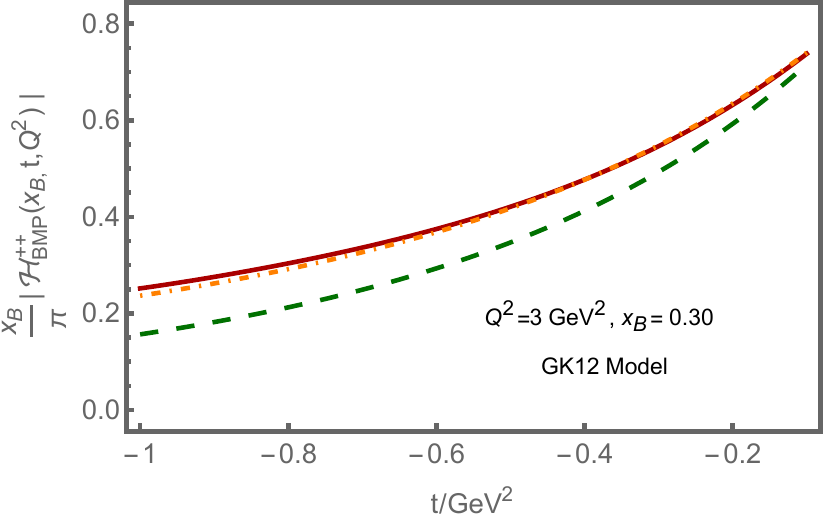}
\includegraphics[width=0.31\textwidth, clip = true]{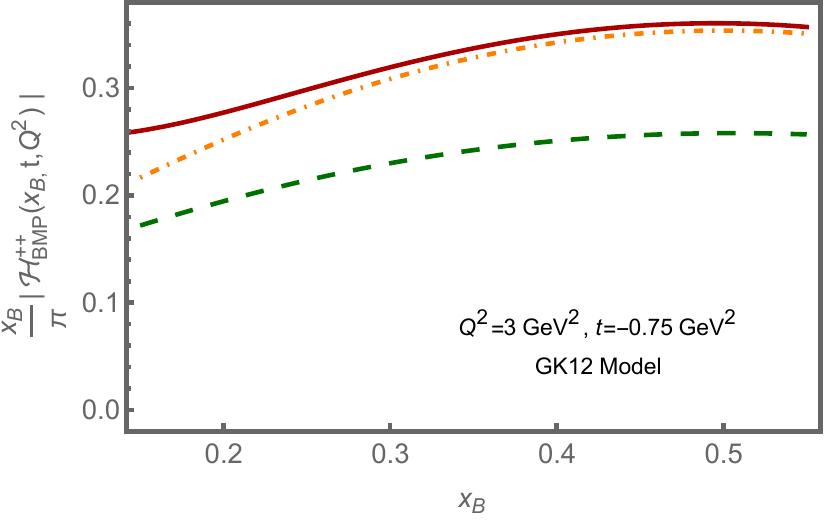}
\\
\includegraphics[width=0.31\textwidth, clip = true]{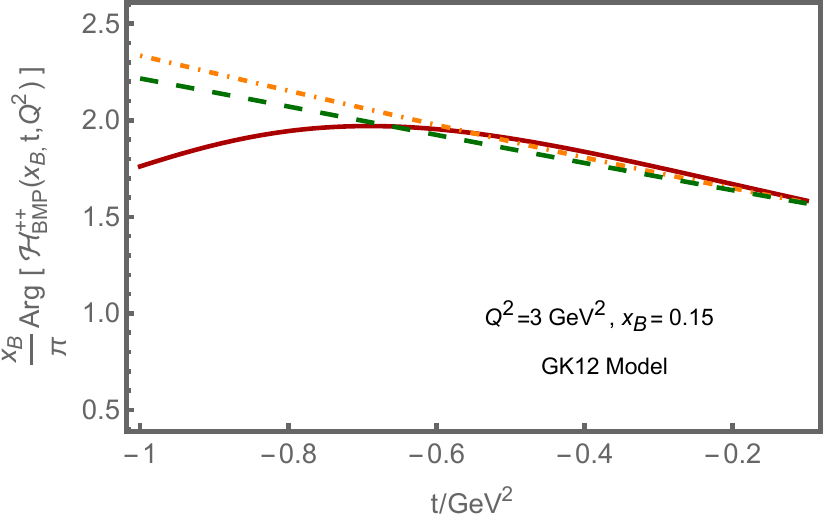}
\includegraphics[width=0.31\textwidth, clip = true]{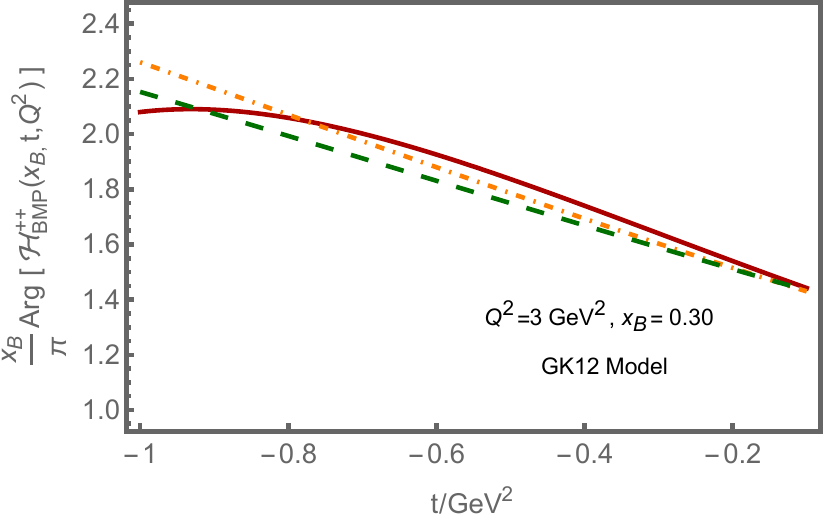}
\includegraphics[width=0.31\textwidth, clip = true]{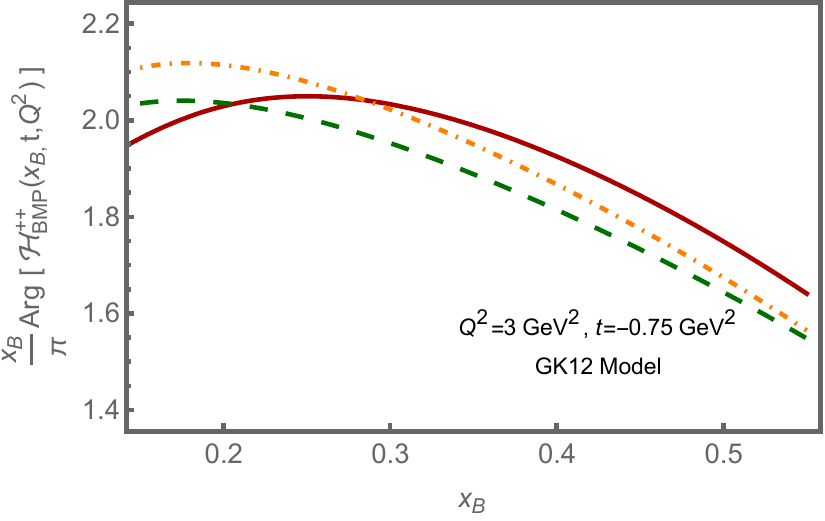}
\caption{Kinematic power corrections to the absolute value and phase of the 
BMP Compton Form Factor $\mathcal{H}^{++}_{\BMP}(x_B,t,Q^2)$.}
\label{H++BMP}
\end{figure*}
\begin{figure*}[ht]
\includegraphics[width=0.31\textwidth, clip = true]{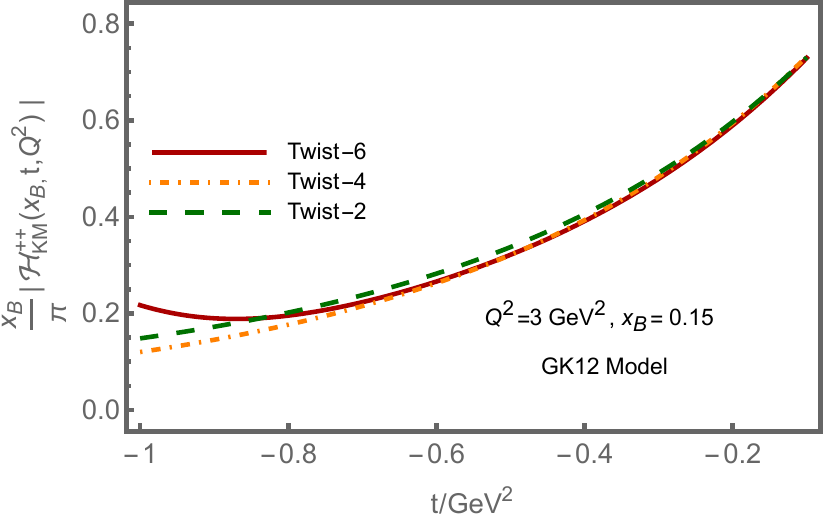}
\includegraphics[width=0.31\textwidth, clip = true]{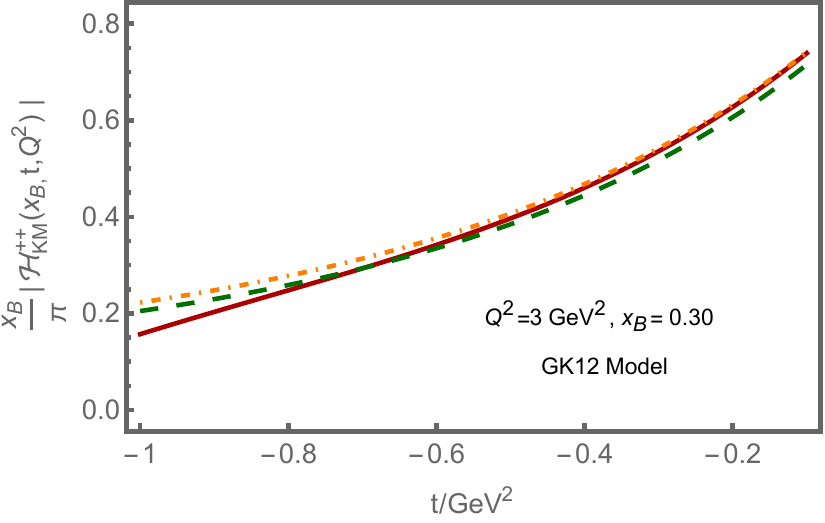}
\includegraphics[width=0.31\textwidth, clip = true]{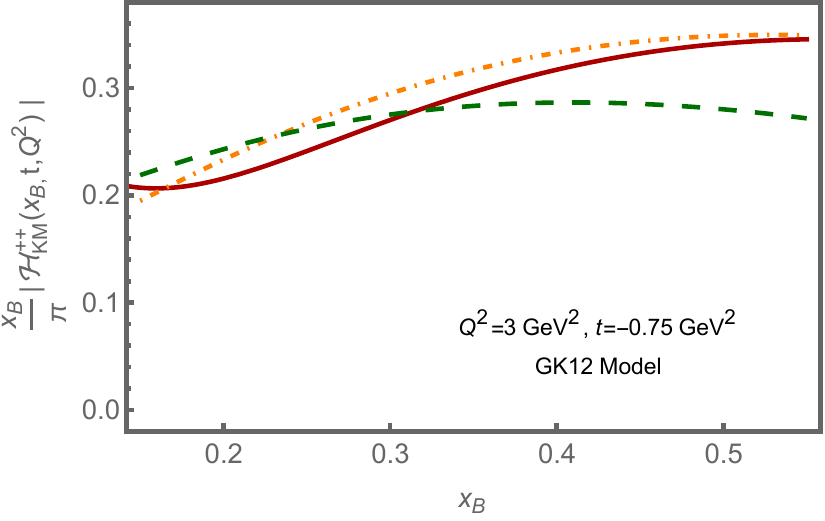}
\\
\includegraphics[width=0.31\textwidth, clip = true]{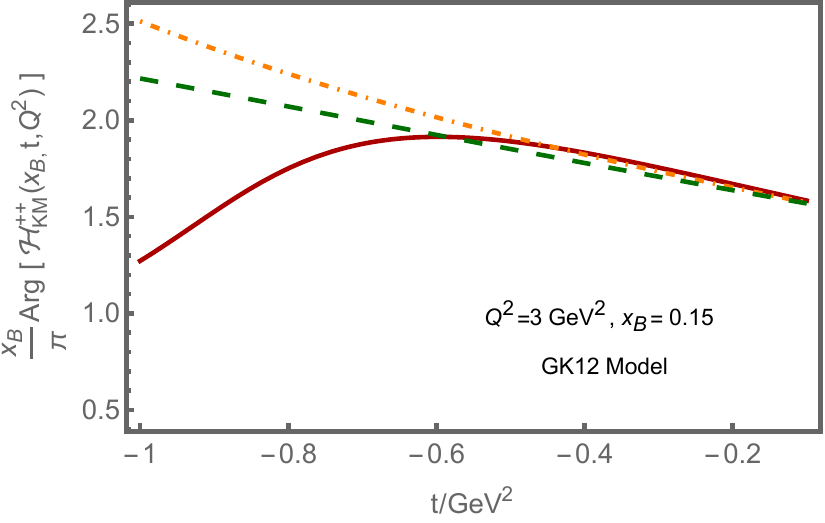}
\includegraphics[width=0.31\textwidth, clip = true]{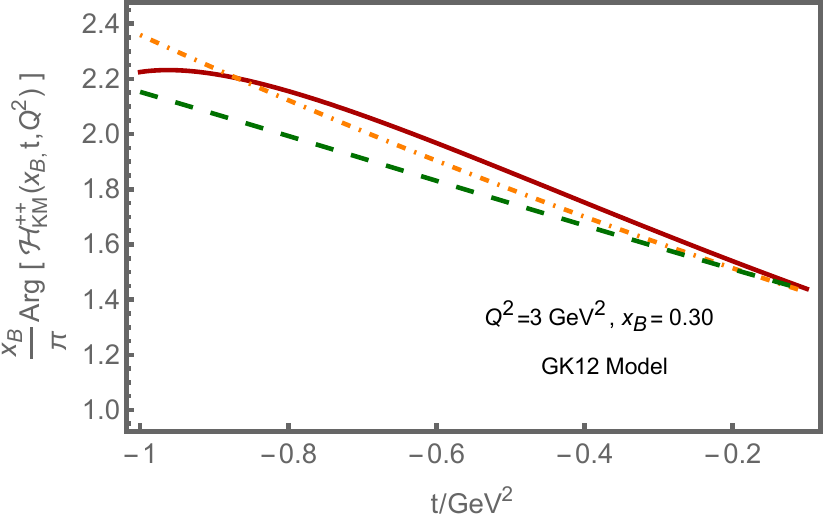}
\includegraphics[width=0.31\textwidth, clip = true]{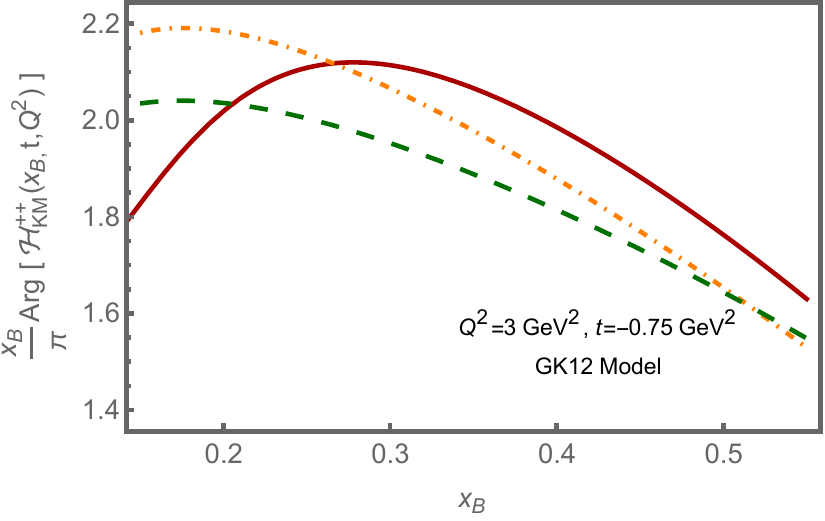}
\caption{Kinematic power corrections to the absolute value and phase of the 
KM Compton Form Factor $\mathcal{H}^{++}_{\BMJ}(x_B,t,Q^2)$.}
\label{H++BMJ}
\end{figure*}


\begin{figure*}[t]
\includegraphics[width=0.31\textwidth, clip = true]{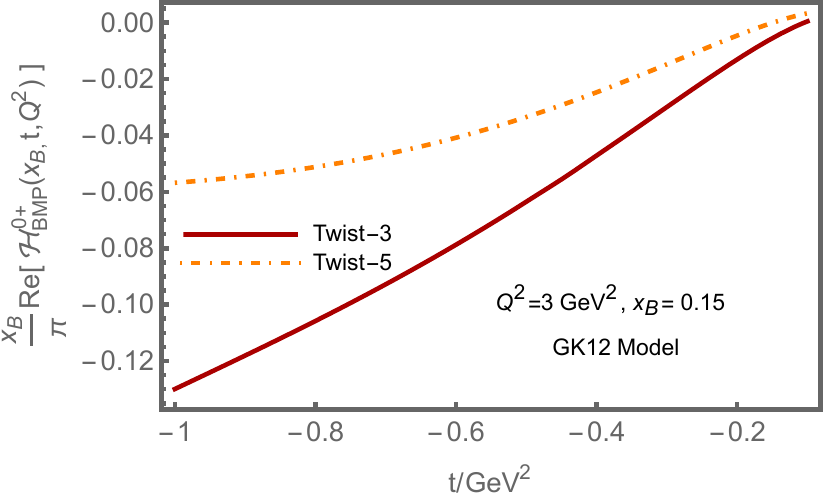}
\includegraphics[width=0.31\textwidth, clip = true]{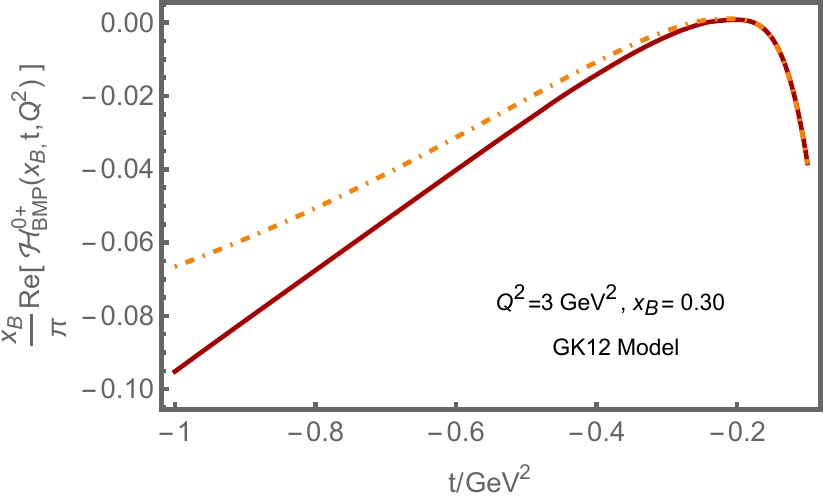}
\includegraphics[width=0.31\textwidth, clip = true]{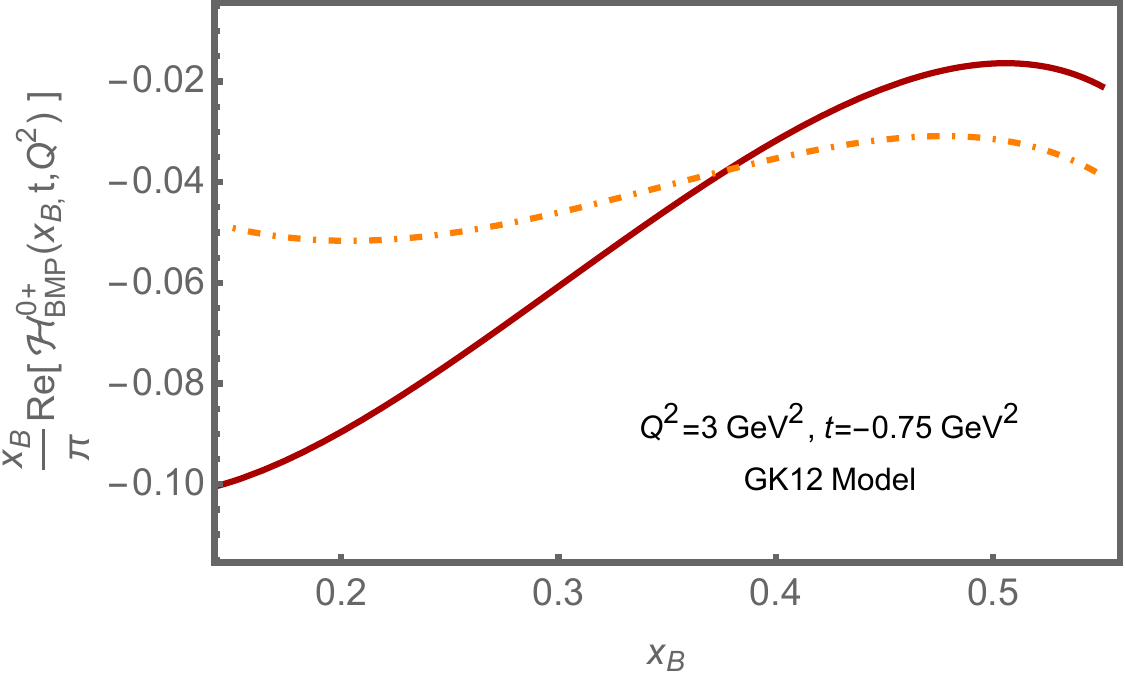}
\\
\includegraphics[width=0.31\textwidth, clip = true]{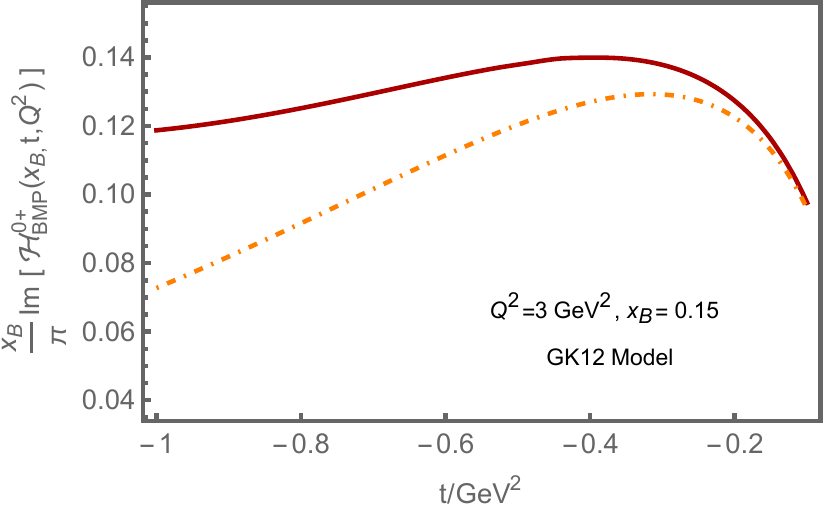}
\includegraphics[width=0.31\textwidth, clip = true]{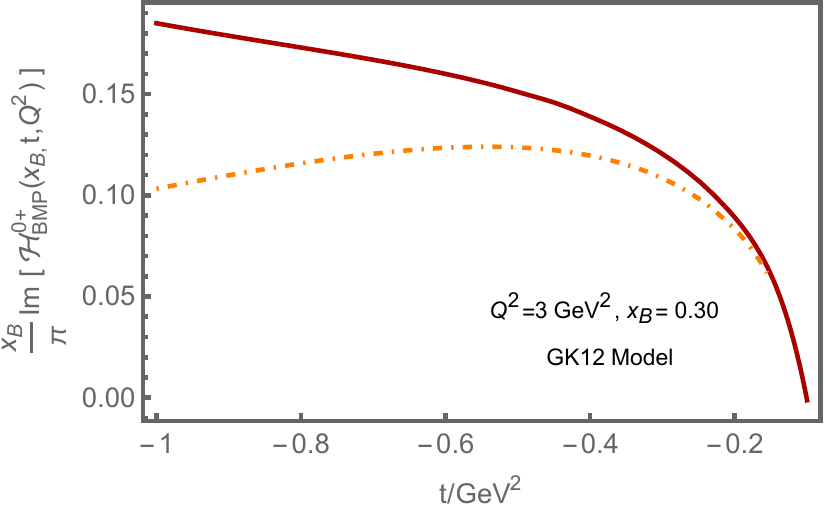}
\includegraphics[width=0.31\textwidth, clip = true]{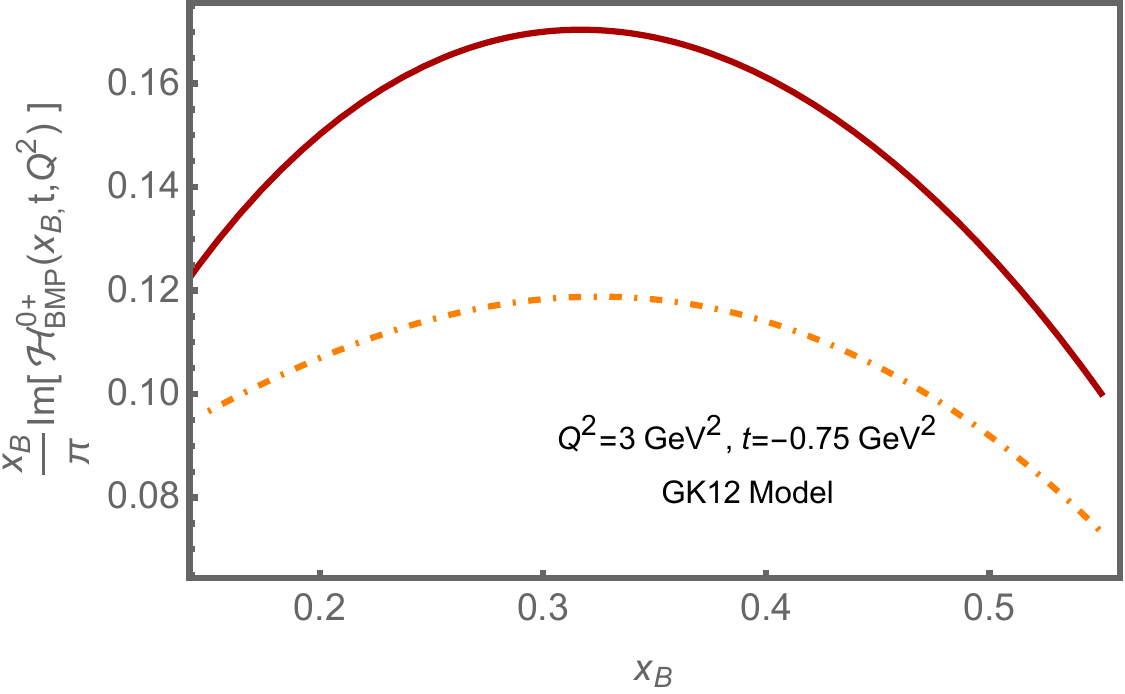}
\caption{Real (upper panels) and imaginary (lower panels) parts of the helicity-flip 
BMP Compton Form Factor $\mathcal{H}^{0+}_{\BMP}(x_B,t,Q^2)$.}
\label{H0+BMP}
\end{figure*}
\begin{figure*}[ht]
\includegraphics[width=0.31\textwidth, clip = true]{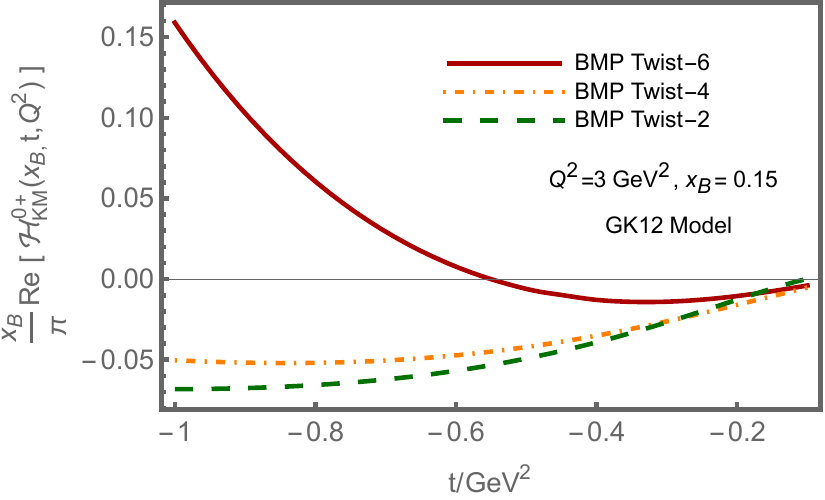}
\includegraphics[width=0.31\textwidth, clip = true]{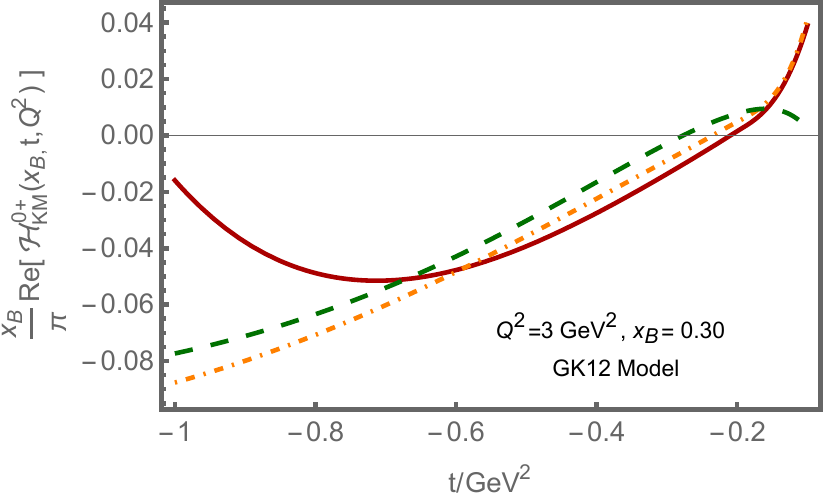}
\includegraphics[width=0.31\textwidth, clip = true]{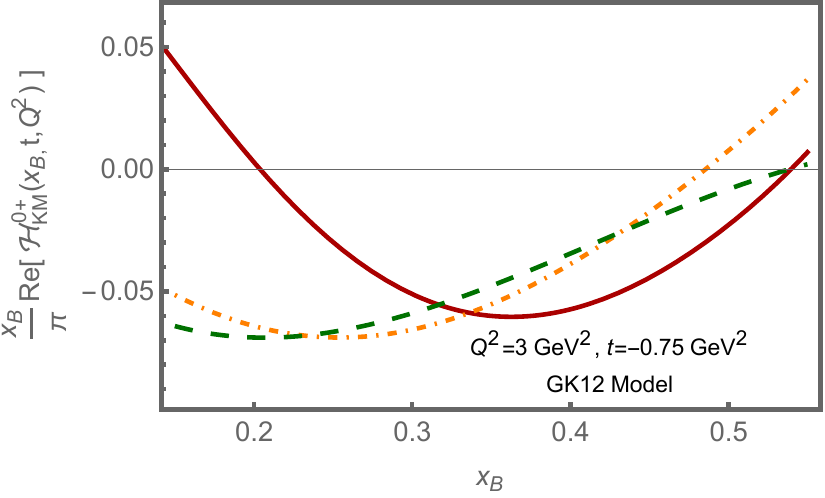}
\\
\includegraphics[width=0.31\textwidth, clip = true]{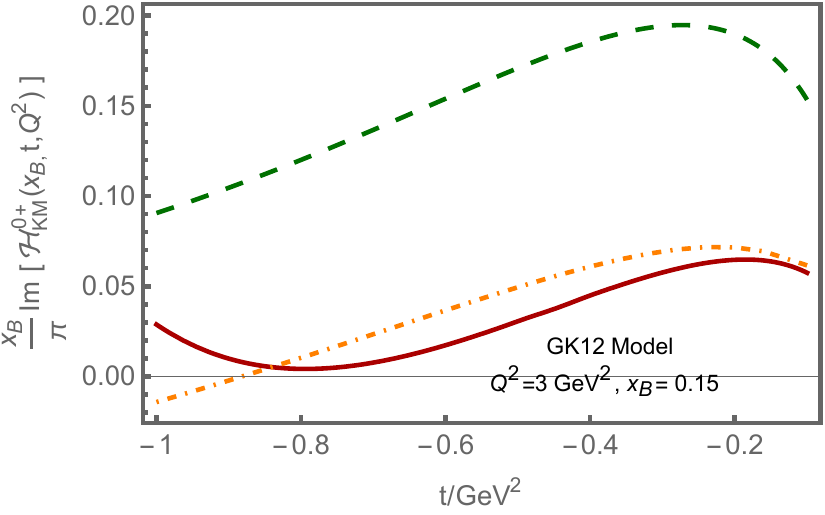}
\includegraphics[width=0.31\textwidth, clip = true]{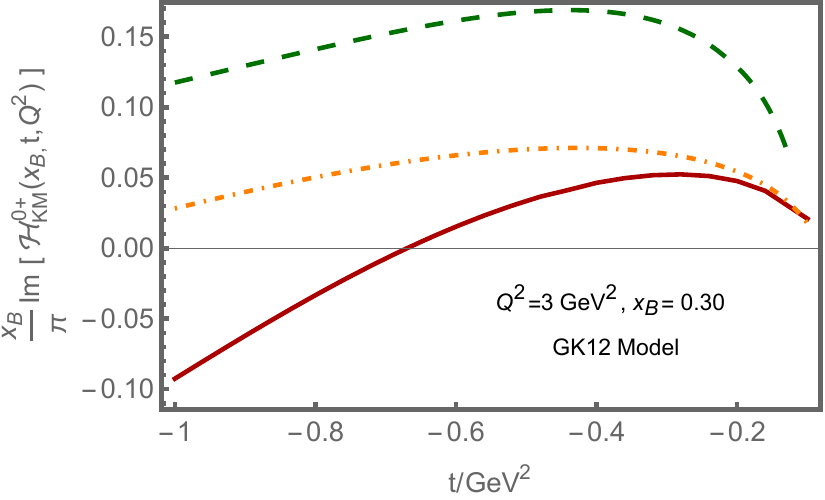}
\includegraphics[width=0.31\textwidth, clip = true]{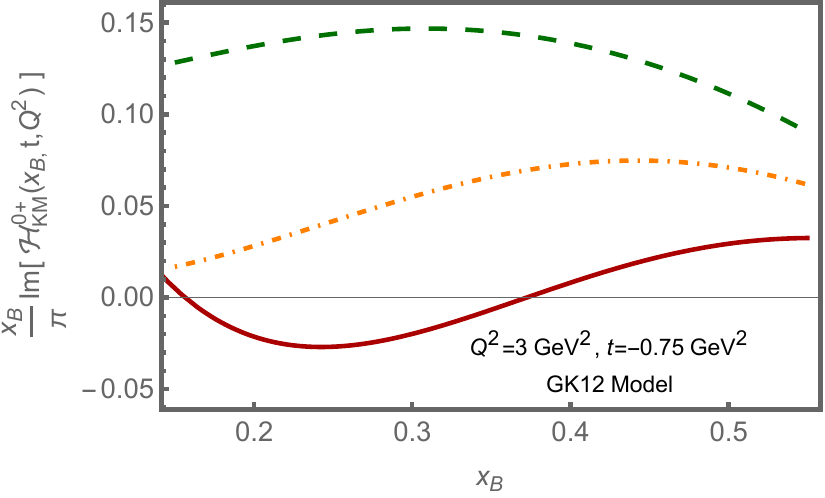}
\caption{Real (upper panels) and imaginary (lower panels) parts of the helicity-flip
KM Compton Form Factor $\mathcal{H}^{0+}_{\BMJ}(x_B,t,Q^2)$.}
\label{H0+BMJ}
\end{figure*}

Note that the power counting $ {\mathcal F}^{++}=\mathcal{O}(1/Q^0)$, ${\mathcal F}^{0+} = \mathcal{O}(1/Q)$, and ${\mathcal F}^{-+} = \mathcal{O}(1/Q^{2})$,
remains the same for both, BMP and BMJ, versions. In particular the difference between the leading, helicity conserving amplitudes, is a higher-twist effect. 
Numerically, however, the difference can be significant since the kinematic factors $\varkappa, \varkappa_0$ are rather large in the experimentally relevant 
kinematics, notwithstanding that they are formally power-suppressed.  
The numerical results presented below are obtained by using the set of BMP CFFs including kinematic power corrections to twist-six accuracy, section~\ref{sec:results},
transforming them to the BMJ CFF basis~\eqref{BMJ-BMP}, and calculating DVCS observables using the expressions provided in Ref.~\cite{Belitsky:2012ch}. 


                            \section{DVCS observables}


To evaluate observables, we need to express the electroproduction cross sections in terms of the BMP CFFs 
$\mathcal F_{\BMP}$. Instead of a direct calculation, we follow the procedure used in Ref.~\cite{Braun:2014sta}, 
transforming BMP CFFs to the KM (BMJ) basis, $\mathcal F_{\BMP}\to \mathcal F_{\BMJ}$, and making use
of the results from Ref.~\cite{Belitsky:2012ch}. This transformation is straightforward and can be thought of as, 
loosely speaking, a Lorentz transformation to a different reference frame.


\begin{figure*}[t]
\includegraphics[width=0.31\textwidth, clip = true]{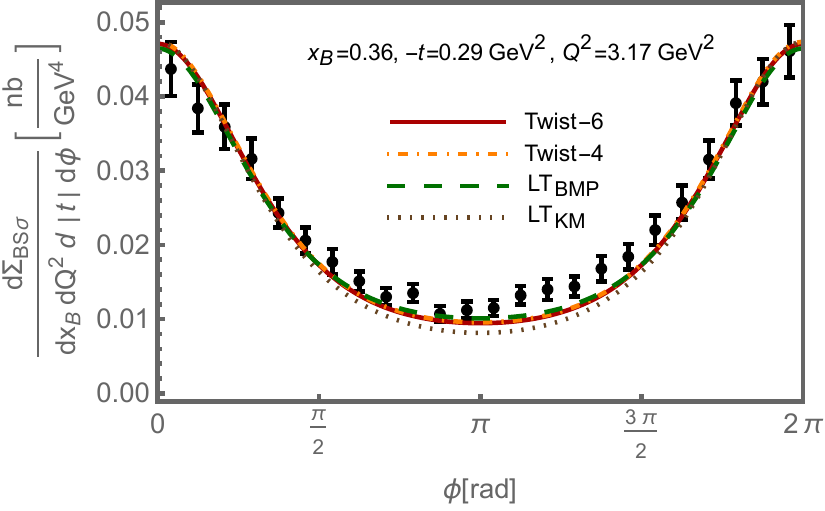}
\includegraphics[width=0.31\textwidth, clip = true]{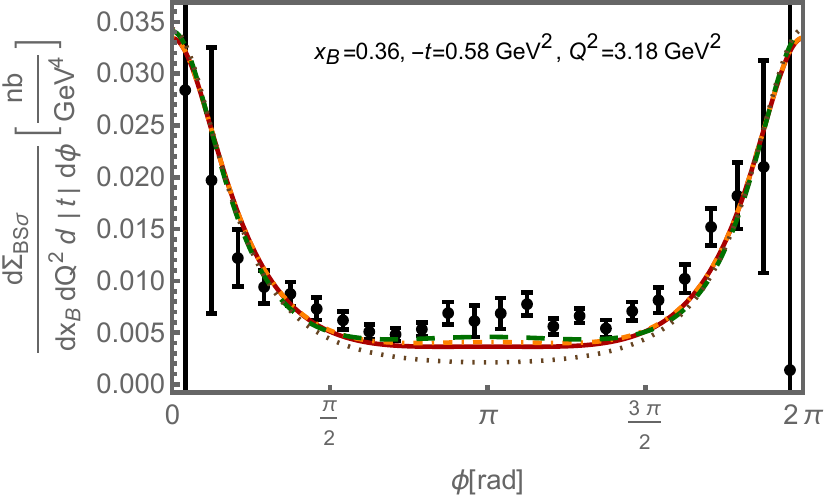}
\includegraphics[width=0.31\textwidth, clip = true]{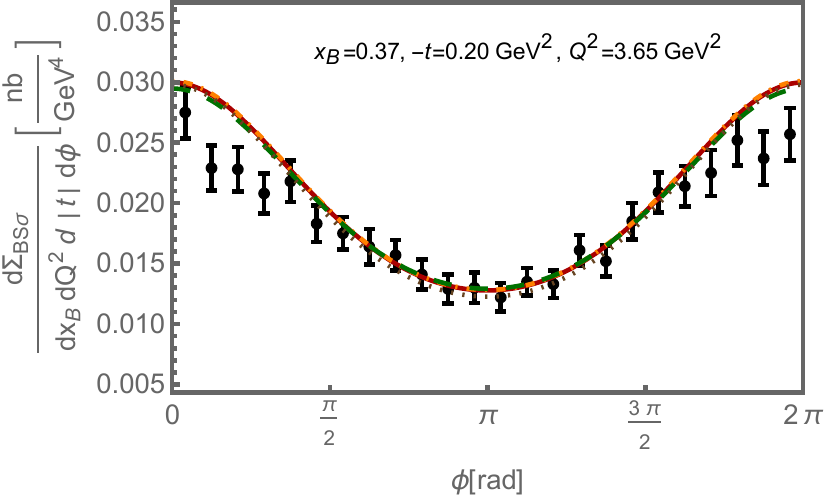}
\\
\includegraphics[width=0.31\textwidth, clip = true]{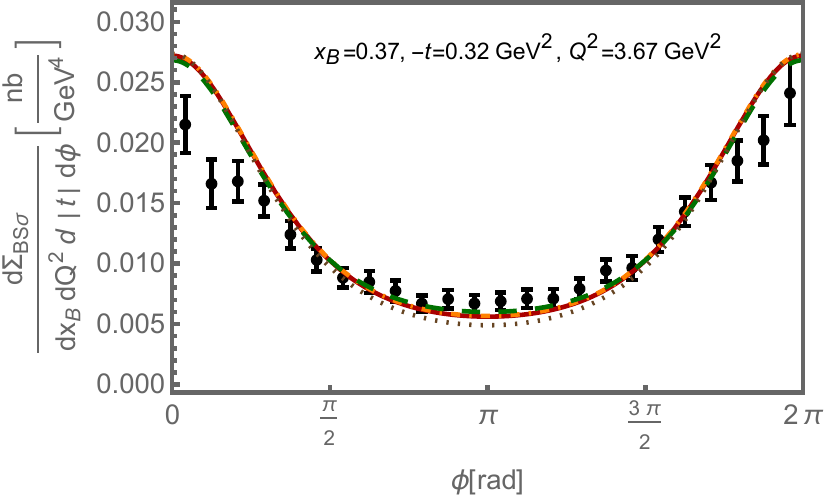}
\includegraphics[width=0.31\textwidth, clip = true]{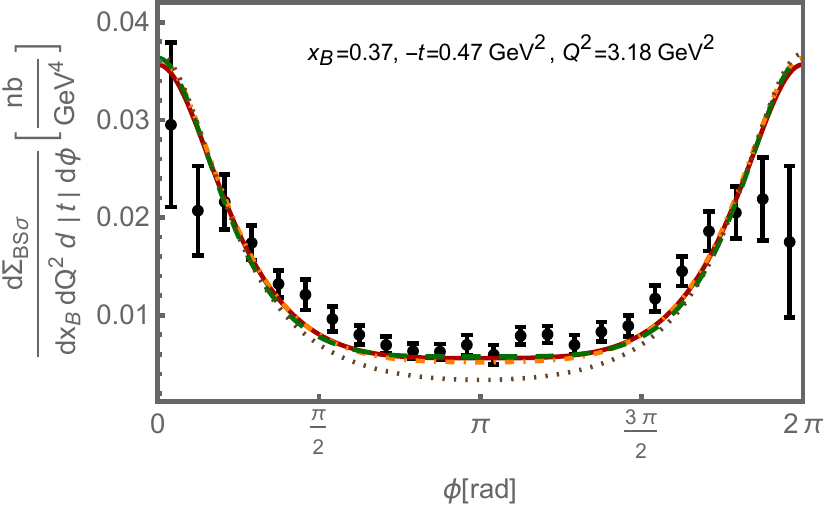}
\includegraphics[width=0.31\textwidth, clip = true]{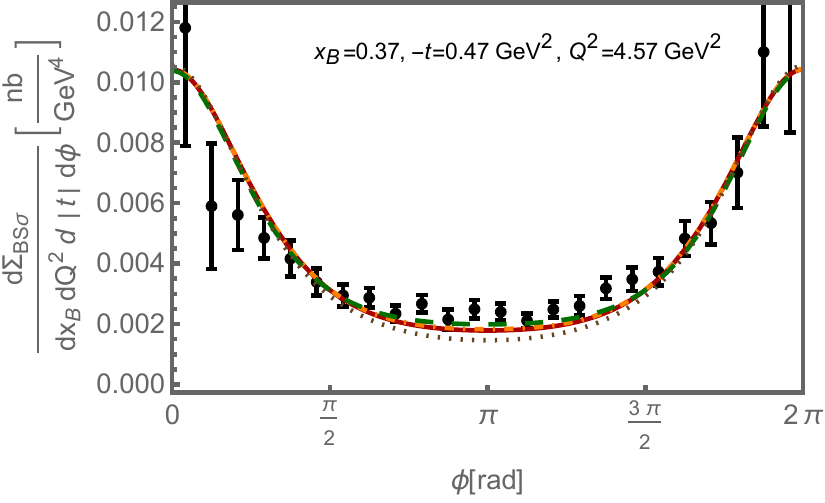}
\\
\includegraphics[width=0.31\textwidth, clip = true]{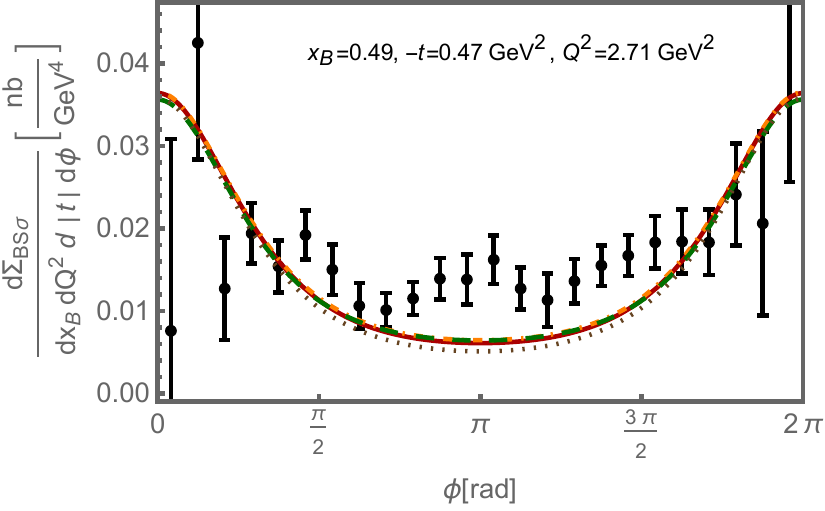}
\includegraphics[width=0.31\textwidth, clip = true]{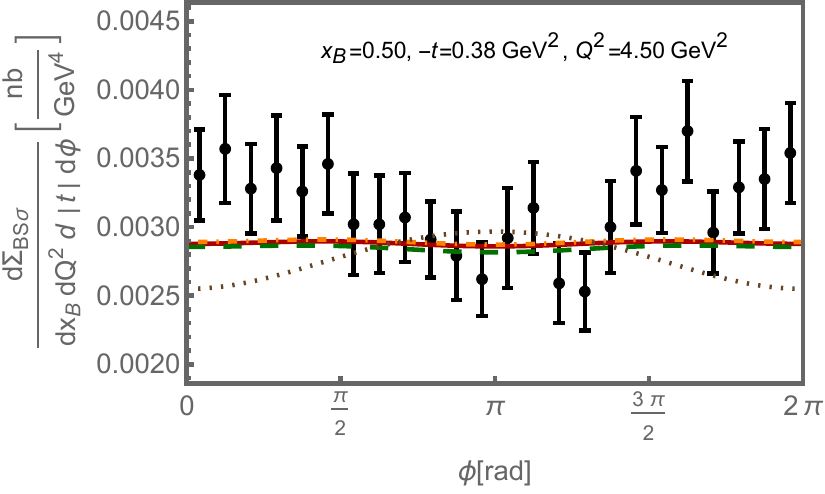}
\phantom{\includegraphics[width=0.31\textwidth, clip = true]{PLOTS/Xsections/BSS_050_038_450}}
\caption{Spin-averaged cross sections from Jefferson Lab HallA~\cite{JeffersonLabHallA:2022pnx}
(selected data sets).}
\label{HallA2022}
\end{figure*}

The results of a numerical calculation presented below are obtained using the GPD model GK12  
Goloskokov and Kroll \cite{Kroll:2012sm}. It is based on the Radyushkin's double distribution ansatz \cite{Radyushkin:1997ki}
and also involves a certain model for an approximate $Q^2$ dependence.
It is convenient for our purposes as all needed $\xi$ derivatives
can be evaluated analytically.

As an example, we consider the helicity-conserving CFF $\mathcal H^{++}(x_B,t,Q^2)$ which gives the 
dominant contribution to the DVCS cross section for the unpolarized target.
The results for the absolute value and phase of $\mathcal H_{\BMP}^{++}(x_B,t,Q^2)$
and $\mathcal H_{\BMJ}^{++}(x_B,t,Q^2)$ are shown in Fig.~\ref{H++BMP} and Fig.~\ref{H++BMJ},
respectively. We choose $Q^2=3\,\text{GeV}^2$ and present, in both cases, the results of the calculation with and without 
power corrections as functions of $t$ for fixed value $x_B=0.15$ (left panels),  $x_B=0.30$ (middle), and, alternatively,
as functions of $x_B$ for fixed $t =-0.75\,\text{GeV}^2$ (right panels). 
As already mentioned, in this work we define the twist expansion as power counting in 
$(qq') = -1/2(Q^2+t)$ which implies that a part of the $1/Q^4$ corrections
is included already in the twist-4 term. The remaining twist-6 contributions remain
small up to $|t|/Q^2 \sim 1/4$ but increase rapidly for larger momentum transfers. 

As another example, we consider the higher-twist helicity-flip CFF $\mathcal H^{0+}(x_B,t,Q^2)$
with a longitudinal virtual photon in the initial state. This CFF 
starts at twist-3 level, $\mathcal H^{0+}\sim 1/Q$. 
The corresponding (kinematic) contribution is traditionally referred to as the Wandzura-Wilczek 
(WW) approximation. The new contribution of this work is the calculation of the 
subleading power twist-5 correction $\sim 1/Q^3$. 
The results in the BMP reference frame are shown in Fig.~\ref{H0+BMP}.
The twist-5 contributions are significant and become of the same order as 
the WW twist-3 term already at $|t|/Q^2\sim 0.3$.  
The corresponding KM CFFs $\mathcal H^{0+}_{\BMJ}$ obtained from 
$\mathcal H^{0+}_{\BMP}$ using the relation in Eq.~\eqref{BMJ-BMP}
are shown in  Fig.~\ref{H0+BMJ}. By this transformation, contributions of different twists 
get mixed. For example, the WW twist-3 contribution in the KM frame is obtained as
a sum of the twist-3 contribution in the BMP frame and the BMP twist-2 
CFF decorated by a kinematic factor $\varkappa_0 \sim \sqrt{-t}/Q$, 
$\mathcal H^{0+}_{\BMJ} = - \mathcal H^{0+}_{\BMP} + \varkappa_0 \mathcal H^{++}_{\BMP} +\ldots$
where the ellipses stand for the terms $\sim 1/Q^3$ and higher powers. 
These two contributions tend to have an opposite sign so that a larger $\mathcal H^{0+}_{\BMP}$ generally leads to 
a smaller $\mathcal H^{0+}_{\BMJ}$, see Fig.~\ref{H0+BMJ}.
In these plots we do not perform a systematic power expansion, but show instead the 
results for  $\mathcal H^{0+}_{\BMP}$ calculated using Eq.~\eqref{BMJ-BMP} literally, 
with a certain approximation for the BMP CFFs $\mathcal H^{++}_{\BMP}$, $\mathcal H^{0+}_{\BMP}$, $\mathcal H^{-+}_{\BMP}$ as inputs.
The dashed curves are obtained by using the leading-twist approximation for  $\mathcal H^{++}_{\BMP}$
and putting the other two BMP CFFs to zero; the dash-dotted curves are obtained using all three BMP CFFs to twist-4 accuracy, 
and the solid curves show the final results including the twist-5 and twist-6 terms. 
The higher-twist corrections are large in all cases and have a nontrivial pattern. However, the helicity flip CFF is small 
in comparison to the helicity-conserving one, cf.~Fig.~\ref{H++BMJ}, so that the results for the cross sections and 
spin asymmetries are not affected strongly.

In the last years new experimental data on DVCS from Jefferson Lab have appeared,
with extended $Q^2$  and $x_B$ phase space reached including considerably smaller
statistical uncertainty as compared to previous results.
In Fig.~\ref{HallA2022} we compare our results on the kinematic power corrections to the spin-averaged
DVCS cross section with the Hall A results \cite{JeffersonLabHallA:2022pnx}. As above, we use the GK12 GPD model
as an example, and take into account Bethe-Heitler contributions as in Ref.~\cite{Braun:2014sta}.
The dotted and dashed curves are calculated in the leading-twist approximation defined in the KM and 
BMP reference frames, the dash-dotted curve shows the calculation with twist-4 corrections included,
and the solid curve presents the full result to twist-6 accuracy. 
We have chosen for this figure the Hall A data sets with larger $|t|/Q^2$ values where the power corrections are more important,
but avoid the ones with the largest $x_B$ values as the GK12 model was not fitted to this range.
The target mass corrections prove to be negligible for all considered cases.

\begin{figure*}[t]
\includegraphics[width=0.31\textwidth, clip = true]{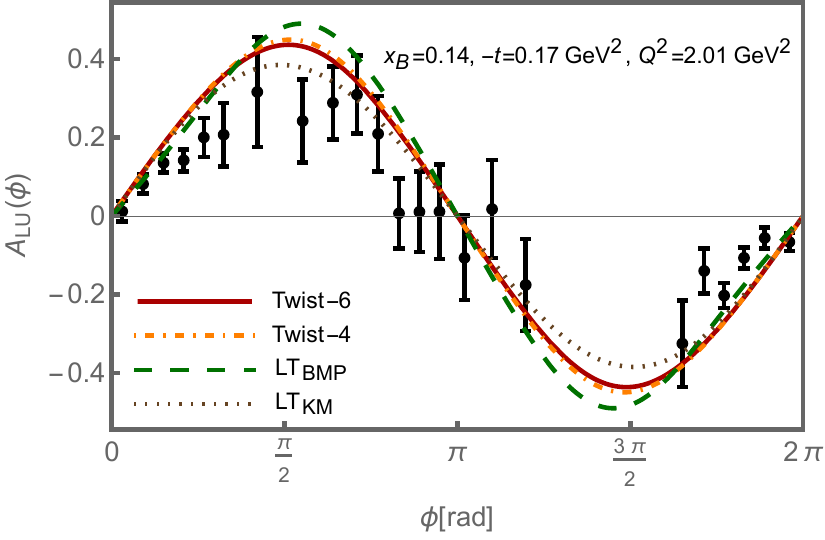}
\includegraphics[width=0.31\textwidth, clip = true]{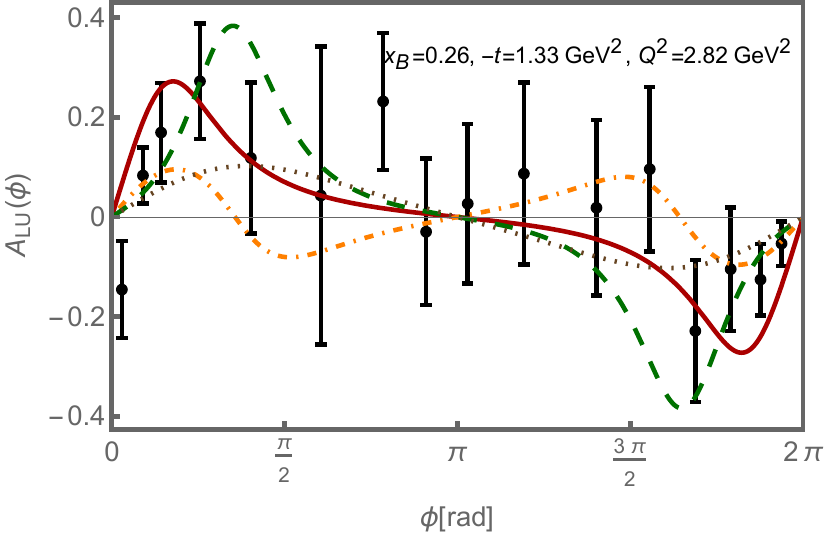}
\includegraphics[width=0.31\textwidth, clip = true]{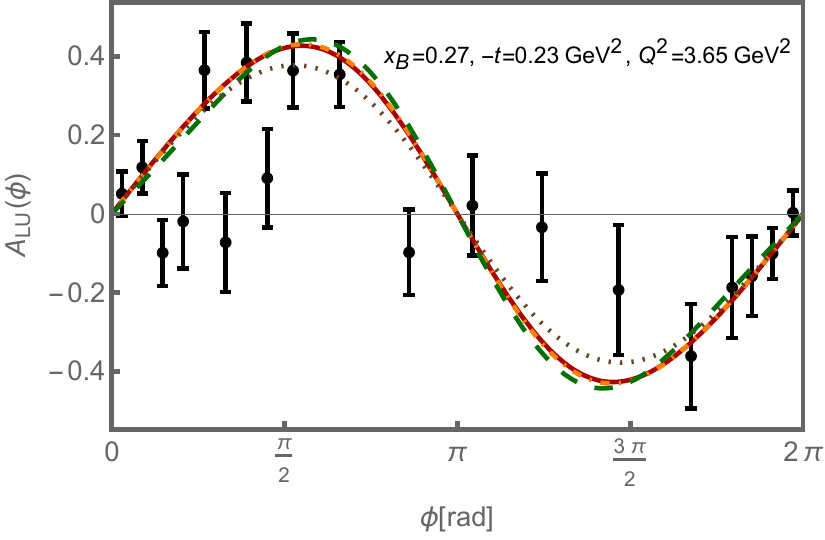}
\\
\includegraphics[width=0.31\textwidth, clip = true]{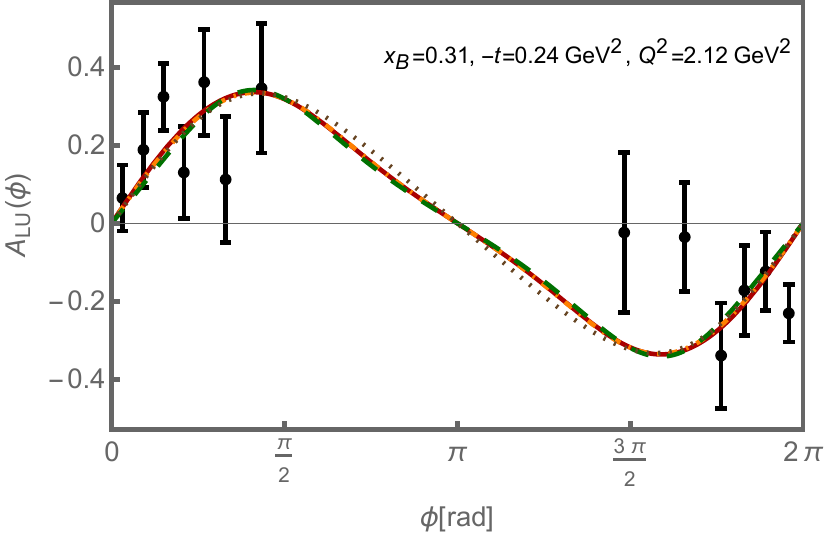}
\includegraphics[width=0.31\textwidth, clip = true]{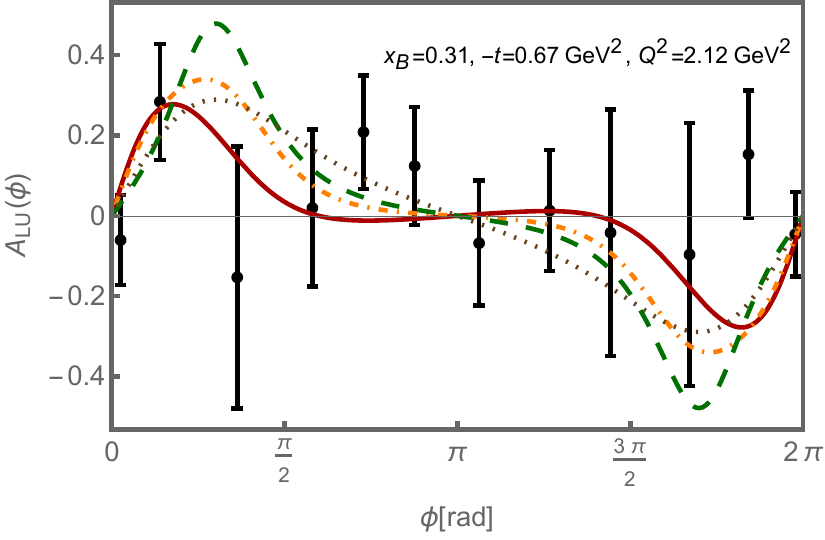}
\includegraphics[width=0.31\textwidth, clip = true]{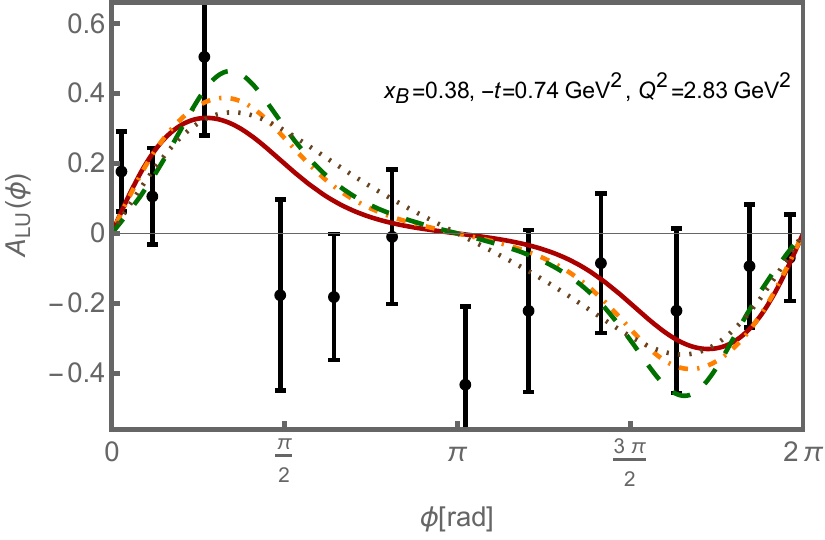}
\\
\includegraphics[width=0.31\textwidth, clip = true]{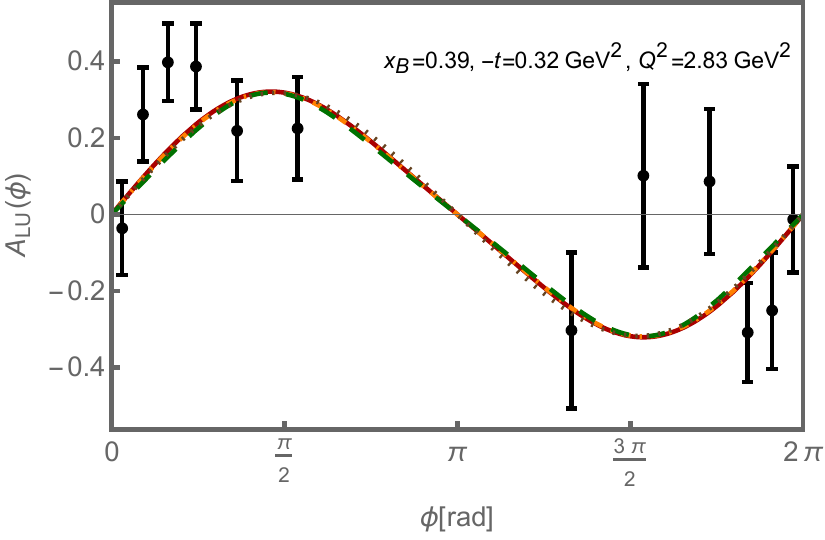}
\includegraphics[width=0.31\textwidth, clip = true]{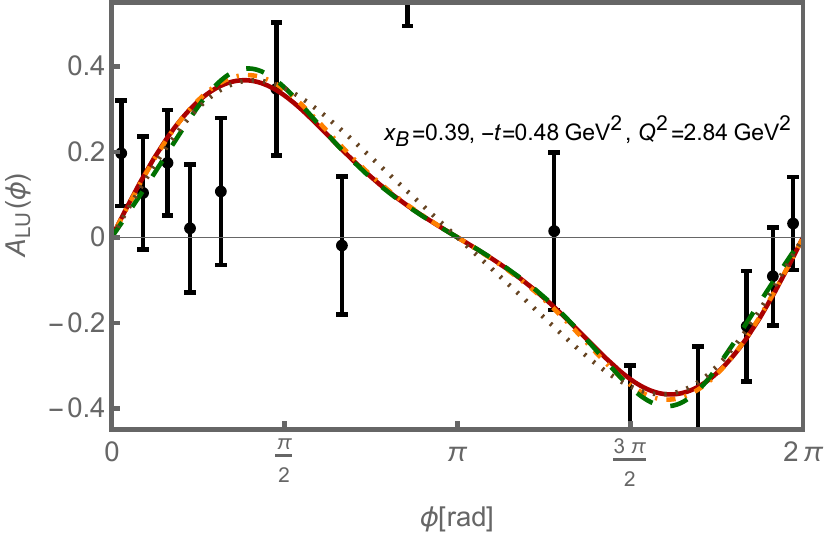}
\includegraphics[width=0.31\textwidth, clip = true]{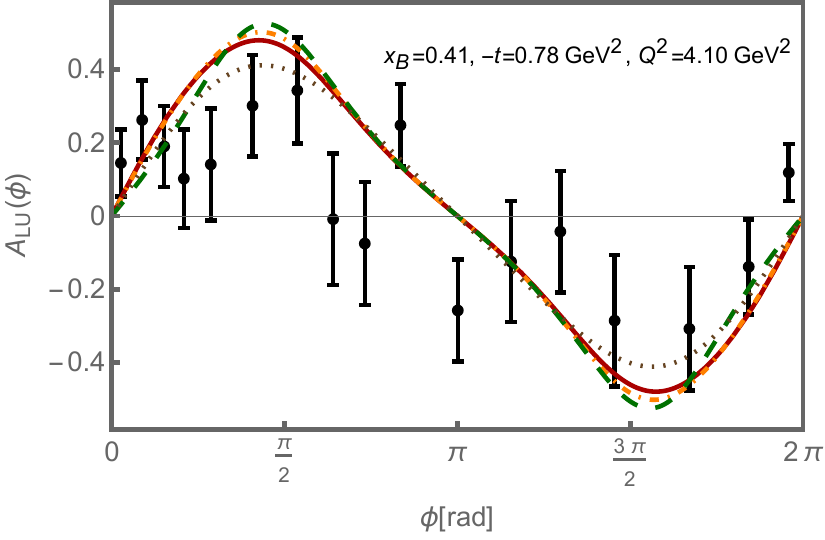}
\caption{Beam spin asymmetries from Jefferson Lab CLAS12  10.6~GeV data set \cite{CLAS:2022syx}
(selected data sets).}
\label{CLAS2022}
\end{figure*}

In Ref.~\cite{CLAS:2022syx} the first measurement of the DVCS
beam-spin asymmetry was reported using the CLAS12 spectrometer with a 10.2 and 10.6 GeV electron beam
scattering off unpolarized protons. Our results in several different approximations 
are compared with their selected data sets in Fig.~\ref{CLAS2022}.
Note that there are several data sets with large values of the momentum transfer $|t|/Q^2 \sim 0.5$
in which case the power corrections become very large, but, in general, their size is moderate
and the overall agreement with the data appears to be quite satisfactory. 
In the future, it would be very interesting to see the spin-averaged cross section measurements 
from CLAS12 in a similar broad $t$-range.


                            \section{Conclusions}


We have studied kinematic power corrections to the DVCS observables including, for the first
time, the contributions of twist-5 and twist-6 to the Compton form factors.  
The motivation for this work is provided by the three-dimensional ``tomographic'' imaging program 
of the proton and light nuclei, with the generalized parton distributions encoding the information on 
the transverse position of quarks and gluons in the proton in dependence on their longitudinal momentum.
The resolving power on the transverse distance is directly limited by the range of the invariant moment transfer $t$ which can be used in the analysis. 
Thus the theoretical control over power corrections $(\sqrt{-t}/Q)^k$ is crucial.

The main thrust of the present calculation has been to find out the range of 
momentum transfers for which the hierarchy of contributions with different power suppression holds, i.e. 
the twist-5,6 contributions are still smaller than twist-3,4 ones.  Our results 
suggest that for $|t|/Q^2 \lesssim 1/4$  the twist expansion is converging for most observables, confirming 
the previous estimate from Ref.~\cite{Braun:2014paa} that was based on the hierarchy of the leading twist-2
and twist-3,4 terms. However, $1/(qq') = 2/(Q^2+t)$ appears to be a better expansion parameter as
compared to the nominal hard scale $1/Q^2$.

Apart from that, we present additional evidence that target mass corrections  $\sim (m/Q)^k$ do not spoil QCD factorization for coherent DVCS on nuclei.
The reason is that such contributions to Compton form factors always involve
additional factors of the skewness parameter, so that the expansion goes in powers 
of $\xi^2 m^2/Q^2$ rather than $m^2/Q^2$.  
This feature was already observed in Ref.~\cite{Braun:2014paa} to twist-4 accuracy, and is now confirmed up to twist-6.
For nuclear targets, effectively, $m\to A m$ and $\xi \to \xi/A$, 
so that the target mass corrections remain essentially the same as for the nucleon and 
are small, apart from the large $x_B$ region.


\begin{acknowledgments}
We thank K. Kumericki for a discussion of the GPD models. 
This work was supported by the Research Unit FOR2926 
funded by the Deutsche Forschungsgemeinschaft 
(DFG, German Research Foundation) under grant No. 409651613.
Y.~J. acknowledges the support of the University Development Fund (UDF) of The Chinese
University of Hong Kong, Shenzhen, under grant No. UDF01003869, and the DFG through
the Sino-German Collaborative Research Center TRR110  
(DFG Project-ID 196253076, NSFC Grant No. 12070131001, - TRR 110).
\end{acknowledgments}

\newpage

\appendix


\section{Helicity amplitudes in the Double Distribution representation}\label{app:DD}


At the intermediate stages of the calculation the following ``double distribution'' (DD) \cite{Radyushkin:1997ki} parametrization 
of the nucleon matrix elements of light-ray vector- and axial-vector operators proves to be the most convenient:
\begin{align}
\mathscr O_V(z_1n,z_2n) &=
\iint dy dz e^{iy P_+z_{12}+i\frac12 \Delta_+ (z_1+z_2-z_{12} z)}
\notag\\&\quad
\times
\biggl\{v_+ \boldsymbol h_-(y,z,t)+
\frac{i(vP)}{z_{12}m^2} \boldsymbol \Phi_+(y,z,t)
\biggr\},
\notag\\
\mathscr O_A(z_1n,z_2n) &=
\iint dy dz e^{iy P_+z_{12}+i\frac12 \Delta_+ (z_1+z_2-z_{12} z)}
\notag\\&\quad
\times\biggl\{a_+  \widetilde{\boldsymbol h}_+(y,z,t)+
\frac{i(a \Delta)}{2z_{12}m^2} \widetilde {\boldsymbol \Phi}_-(y,z,t)
\biggr\}.
\end{align}
The variables $x,y$ are related to the original Radyushkin's notation
as $y\equiv\beta$ and $z\equiv\alpha$, and the integration goes over the region $|y|+ |z|\leq 1$. 
The subscripts $\pm$ of the DDs (boldface) indicate parity under the $(y,z)\mapsto (-y,-z)$ transformation, namely
\begin{align}
\boldsymbol h_-(y,z,t) &= -\boldsymbol h_-(-y,-z,t)\,,  
\notag\\
\widetilde{\boldsymbol h}_+(y,z,t) &= \widetilde {\boldsymbol h}_+(-y,-z,t)\,,
\end{align}
etc. 
The DD $\widetilde {\boldsymbol \Phi}_-$ is an odd function of $z$ whereas all other DDs are even functions of $z$. 
The DDs $\boldsymbol\Phi_+$ and $\widetilde{\boldsymbol\Phi}_-$ can alternatively be written as~\cite{Teryaev:2001qm}
\begin{align}
\boldsymbol\Phi_+(y,z)&=\partial_y \boldsymbol f(y,z) +\partial_z\boldsymbol g(y,z)\,,
\notag\\
\widetilde{\boldsymbol\Phi}_-(y,z)&=\partial_y \widetilde{\boldsymbol f}(y,z) +\partial_z \widetilde{\boldsymbol g}(y,z)\,.
\end{align}
The ``standard'' GPDs $H,~E,~\widetilde H,~\widetilde E$ \eqref{defGPD} can be expressed 
in terms of the DDs defined above as follows~\cite{Braun:2014paa},
\begin{align}
(H\!+\!E)(x,\xi,t) & =\iint dy\, dz\,\delta(x- y -\xi z)\,\boldsymbol h_-(y,z,t)\,,
\notag\\
-E(x,\xi,t) &=\iint dy dz\,\delta(x-y-\xi z)\,
\notag\\&\qquad
\times\Big(\boldsymbol f(y,z,t) +\xi \boldsymbol g (y,z,t)\Big),
\notag\\
\widetilde H(x,\xi,t) & =\iint dy\, dz\,\delta(x- y -\xi z)\,\widetilde{\boldsymbol h}_+(y,z,t)\,,
\notag\\
-\widetilde E(x,\xi,t) &=\frac1\xi\iint dy dz\,\delta(x-y-\xi z)\,
\notag\\&\qquad
\times \Big(\widetilde{\boldsymbol f}(y,z,t)
+\xi \widetilde{\boldsymbol g} (y,z,t)\Big).
\end{align}

The calculation follows the same routine as for a scalar target \cite{Braun:2022qly}, but is more 
cumbersome due to a proliferation of Lorentz structures. In the expressions given below 
$\boldsymbol h_- \equiv \boldsymbol h_-(y,z)$, $\widetilde{\boldsymbol  h}_+ \equiv \widetilde{\boldsymbol  h}_+(y,z)$, etc.   
We use rescaled variables $\t$, $\mm$, $\Pperp$ defined in Eq.~\eqref{rescale} and the notation
\begin{align}
\Dw \equiv y\, \partial_w\,,
\end{align}
where
\begin{align}
w\equiv w(y,z)&=\frac12\left(\frac y\xi + z+1\right)\,, 
\notag\\
 w(-y,-z)&=1-w(y,z)\,.
\end{align}  
The coefficient functions are defined in Eq.~\eqref{Tfunctions}. 

We write the result for the helicity-conserving amplitude $\mathcal A_{A,V}^{\pm\pm}$ as a sum of four terms
following the notation in Eq.~\eqref{Apmpm}.
We obtain
\allowdisplaybreaks{
\begin{widetext}
\begin{subequations}
\begin{align}
V^{(1)}_0 & = 
\iint dy dz\,\boldsymbol h_-  \biggl\{- \left(1+\frac{\t}{4 }\right) T_0(w)
-  \frac{\t}{2} T_{10}(w)
+ \biggl[\t\Big(1+\frac1\xi \Dw\Big)-\frac12 \Pperp \Dw^2\biggr] \Big[T_2(w) + 2\t\,T_V(w)\Big] 
\notag\\&\quad
+ \frac{\t^2}4 \,T_{11}(w)
+\biggl[ \frac32 \t \Big(1+\frac1{2\xi} \Dw\Big)\Big(\frac \t \xi -\Pperp \Dw\Big) \Dw
+\frac18 |\widehat{P}_\perp|^4 \Dw^4 \biggr]  T_3(w)
\biggr\},
\\
V^{(2)}_0 & = 
\iint dy dz\, \boldsymbol \Phi_+\, \biggl\{
\Big(1+\frac{\t}{4 }\Big)T_1(w)
+ \frac{\t}{2} \, \Li_2(w)
+\frac12 y \Big(\frac \t\xi -\Pperp \Dw\Big) T_2(w)
+ \frac{\t^2}{4} \Big(\Li_2(w)-2\bar w\ln\bar w\Big)
\notag\\&\quad
+ y \t\Big(\frac \t \xi -\Pperp \Dw\Big) T_V(w)
- \frac12 y \biggl[-\frac{\t^2}{2\xi^2} + \t {\Pperp} \Big(1+\frac1\xi \Dw\Big)-\frac14 |\widehat{P}_\perp|^4 \Dw^2\biggr] \Dw  T_3(w)
\biggr\}
\notag\\&\quad
- \mm  \iint dy dz \,\boldsymbol h_-
\biggl\{ {\Dw\big(T_2(w) +  2\,\t\, \mathrm T_V(w)\big)}
{{-}} \biggl[\t\Big(1+\frac1\xi \Dw\Big) -\frac12 \Pperp \Dw^2\biggr]  T_3(w)\biggr\}
\end{align} 
\end{subequations}
%
%
\begin{subequations}
\begin{align}
   A_0^{(1)} & =  
  \iint dy dz\,\widetilde{\boldsymbol h}_+ \biggl\{
  \Big(1+\frac{\t}{4}\Big) T_0(w)
+ \frac{\t}{2}\,T_{10}(w)
-\biggl[\t\Big(1+\frac1\xi \Dw\Big)  -\frac12 \Pperp {\Dw^2}\biggr] \Big[T_2(w) - 2 \t\, T_A(w)\Big]
\notag\\&\quad
+\frac{\t^2}{4} T_{10}(w)
+ \frac92 \biggl[
 \t \Big(1+\frac1{2\xi}\Dw\Big)\Big(\frac{\t}\xi -\Pperp \Dw\Big) + \frac{1}{12} |\widehat{P}_\perp|^4 \Dw^3
\biggr] \Dw \Big[T_A(w) - \frac12 T_{00}(w)
\Big]\biggr\},
\\
{A}_0^{(2)} & = 
  \iint dy dz\,\widetilde{\boldsymbol \Phi}_- \,\biggl\{{-  \Big(1+\frac \t 4\Big) T_1(w)}
- {\frac{\t}{2}}\, \Li_2(w) - {\frac12} y\,\Big(\frac \t\xi -\Pperp \Dw\Big) \Big[T_2(w) - 2 \t\, T_A(w)\Big]
\notag\\&\quad
-{ \frac{\t^2}{4}} \Big(\Li_2(w)-\frac32 w\Big)
 + { \frac34} y \biggl[\frac{\t^2}{\xi^2} - 2 \t\Pperp\Big(1+\frac1\xi \Dw\Big)
+ \frac12 |\widehat{P}_\perp|^4 \Dw^2\biggr] 
\Dw \biggl[T_A(w) - \frac12 T_{00}(w)
\biggr]\biggr\}
\notag\\
&\quad
+ { 2}\mm \iint dy dz\,
\widetilde{\boldsymbol h}_+\,  \biggl\{
\Big(1+\frac1{2\xi}\Dw\Big) \Big[T_2(w)- 2 \t\, T_A(w)\Big] 
 \notag\\  &\qquad\qquad
 - \frac32 {\biggl[ \frac{\t}\xi +\frac{\Pperp}{2\xi}\Dw^2 + 2\Big(\frac{\t}{\xi} -\Pperp \Dw\Big) \Big(1+ \frac{1}{2\xi} \Dw\Big) \biggr]}\Dw
\biggl[ T_A(w) - \frac12 T_{00}(w)
\biggr]\biggr\}.
\end{align}
\end{subequations}
%

Helicity-flip amplitudes $\mathcal A_{A,V}^{0\pm}$ can be written as sum of six invariant functions defined in Eq.~\eqref{A0pm}:
\begin{subequations}
\begin{align}
V_1^{(1)} &= \iint dy dz\,\boldsymbol h_- \biggl\{ 
- T_{10}(w)
 + \frac{\t}{2} T_{11}(w)
+   \biggl[\t\Big(1+\frac1\xi \Dw\Big)-\frac12\Pperp \Dw^2\biggr]T_V(w)\biggr\},
\\
V_1^{(2)} &= \iint dy dz\,\boldsymbol h_-\,\biggl\{
{\Dw}T_{10}(w)
 - \frac{\t}{2} {\Dw}T_{11}(w)
- 3 \biggl[\t\Big(1+\frac1{2\xi} \Dw\Big)-\frac16\Pperp \Dw^2\biggr] \Dw T_V(w)\biggr\},
\\
V_1^{(3)} &= \iint dy dz \, y\, \boldsymbol \Phi_+\,\biggl\{
T_{10}(w)
 - \frac{\t}{2} T_{11}(w)
-  \biggl[\t\Big(1+\frac1\xi \Dw\Big)-\frac12\Pperp \Dw^2\biggr] T_V(w)\biggr\}
\notag\\
 &\quad
 + \mm \iint dy dz\, \boldsymbol h_-\, \Dw^2 T_V(w),
\end{align}
\end{subequations}
%
%
\begin{subequations}
\begin{align}
  A_1^{(1)} &= 
    \iint dy dz\,{\widetilde{\boldsymbol h}}_+ \,\biggl\{ \Big(1+\frac{\t}{2}\Big)T_{10}(w)
+  \biggl[ \t \Big(1+\frac1\xi \Dw\Big) - \frac12 \Pperp \Dw^2\biggr]
 T_{A}(w)\biggr\},
\\
  A_1^{(2)}  &= 
    \iint dy dz\,{\widetilde{\boldsymbol h}}_+ \,\biggl\{  - \Big(1+\frac{\t}{2}\Big) {\Dw}  \, T_{10}(w)
  - 3  \biggl[\t \Big(1+\frac1{2\xi} \Dw\Big)-\frac16 \Pperp \Dw^2\biggr] \Dw  T_{A}(w)\biggr\},
\\
  A_1^{(3)} &= 
    \iint dy dz\,y\, \widetilde{ \boldsymbol \Phi}_{-}\,\biggl\{ 
 - 
 \Big(1+\frac{\t}{2}\Big)   \, T_{10}(w)
 - 
 \biggl[ \t \Big(1+\frac1\xi \Dw\Big) - \frac12 \Pperp \Dw^2\biggr] T_{A}(w)
\biggr\}
 \notag\\ &\quad
  + {4} \mm \iint dy dz\,  
\tilde{\boldsymbol h}_+\, 
\Big(1+\frac1{4\xi} \Dw\Big) \Dw T_{A}(w).
\end{align}
\end{subequations}
Finally, the amplitudes $\mathcal A_{A,V}^{\mp\pm}$ with helicity flip by two units,
can be written as a sum of 6 terms as in  Eq.~\eqref{Amppm}.
We get
%
\begin{subequations}
\begin{align}
V_2^{(1)} &=
\iint dy dz\,\boldsymbol h_-\,\biggl\{  2 \Big(1+\frac{\t}{4}\Big)  \Dw\,T_{11}(w)
+ 3 \left[\t\Big(1+\frac{1}{2\xi} \Dw\Big) - \frac16 \Pperp \Dw^2\right] \Dw T_V(w)\biggr\},
\\
V_2^{(2)} &=
\Pperp \iint dy dz\,\boldsymbol h_-\,\biggl\{  
- \frac12 \Big(1+\frac{\t}{4}\Big) \Dw^2\,T_{11}(w)
- \frac32 \biggl[\t\Big(1+\frac1{3\xi} \Dw\Big) - \frac1 {12}{\Pperp}  \Dw^2\biggr] \Dw^2 T_V(w)\biggr\},
\\
V_2^{(3)} &=
\Pperp \iint dy dz\, y\, \boldsymbol \Phi_+ \,\biggl\{
 - \frac12 \Big(1+\frac{\t}{4}\Big)  \Dw\,T_{11}(w)
 -\frac34 \biggl[\t\Big(1+\frac{1}{2\xi} \Dw\Big) - \frac16 \Pperp \Dw^2\biggr] \Dw T_V(w)\biggr\}
\notag\\ &\quad
+{ \frac{1}{4} }\Pperp \mm \iint dy dz\, \boldsymbol h_-\,   \Dw^3 T_V(w),
\end{align}
\end{subequations}
%
\begin{subequations}
\begin{align}
A_2^{(1)} &=
\iint dy dz\,\widetilde{\boldsymbol h}_+\,\biggl\{ 
2 \Big(1+\frac{\t}{4}\Big) \Dw\,T_{10}(w)
+ 3 \biggl[\t\Big(1+\frac1{2\xi}\Dw\Big)-\frac16\Pperp \Dw^2\biggr]\Dw T_A(w)\biggr\},
\\
A_2^{(2)} &=
\Pperp \iint dy dz\,\widetilde{\boldsymbol h}_+\,\biggl\{ 
 - \frac12 \Big(1+\frac{\t}{4}\Big) \Dw^2 \,T_{10}(w)
 - {\frac32} \biggl[ \t\Big(1+\frac1{3\xi} \Dw\Big)-\frac1{12} \Pperp \Dw^2\biggr] \Dw^2T_A(w)\biggr\},
\\
A_2^{(3)} &=
\Pperp \iint dy dz\,y\,\widetilde{ \boldsymbol \Phi}_-\biggl\{ 
 -{\frac12} \Big(1+\frac{\t}{4}\Big)  \Dw\,T_{10}(w)
 -{\frac34} \biggl[\t\Big(1+\frac1{2\xi}\Dw\Big)-\frac16\Pperp \Dw^2\biggr]\Dw  T_A(w)\biggr\}
\notag\\ &\quad +
{\frac32} \Pperp \mm \iint dy dz\, \widetilde{\boldsymbol  h}_+\,\Big(1 + \frac1{6\xi} \Dw\Big)\Dw^2  T_A(w)\,. 
\end{align}
\end{subequations}

Starting from these expressions, it is straightforward to rewrite the results in terms of GPDs using 
the following identities:
\begin{align}
\iint dy dz \, \boldsymbol \Phi_+(y,z)\,Y(w) & = \frac1{2\xi} \int dx\, Y^\prime\left(\frac{x+\xi}{2\xi}\right) E(x,\xi,t)\,,
\notag\\
\iint dy dz \,\widetilde{\boldsymbol\Phi}_-(y,z) \,Y(w)&= \frac1{2} \int dx\, Y^\prime\left(\frac{x+\xi}{2\xi}\right) \widetilde  E(x,\xi,t)\,,
\notag\\
\iint dy dz \, y\,\boldsymbol \Phi_+(y,z)\,  \Dw^k\, Y(w) & = \left(-2\xi^2\partial_\xi\right)^{k+1}
\frac1{2\xi} \int dx\, Y\left(\frac{x+\xi}{2\xi}\right) E(x,\xi,t)\,,
\notag\\
\iint dy dz \, y\,\widetilde{\boldsymbol \Phi}_-(y,z)\,  \Dw^k \,Y(w) & = \left(-2\xi^2\partial_\xi\right)^{k+1}
\frac1{2} \int dx\, Y\left(\frac{x+\xi}{2\xi}\right) \widetilde E(x,\xi,t)\,,
\notag\\
\iint dy dz \,\boldsymbol h_-(y,z)\,  \Dw^k\, Y(w) & = \left(-2\xi^2\partial_\xi\right)^{k}
 \int dx\, Y\left(\frac{x+\xi}{2\xi}\right)\Big( H(x,\xi,t) + E(x,\xi,t) \Big)\,,
\notag\\
\iint dy dz \, \widetilde{\boldsymbol h}_+(y,z)\,  \Dw^k \,Y(w) & = \left(-2\xi^2\partial_\xi\right)^{k}
 \int dx\, Y\left(\frac{x+\xi}{2\xi}\right) \widetilde H(x,\xi,t)\,,
\end{align}
where $Y(w)$ is an arbitrary function.
The final results in the GPD representation are given in section~\ref{sec:results} .
\end{widetext}

\bibliography{references}

\end{document}